\documentclass[11pt]{article}
\usepackage[utf8]{inputenc}
\usepackage[frozencache, cachedir=.]{minted}
\usepackage{amsmath}
\usepackage{amsfonts}
\usepackage[utf8]{inputenc}
\usepackage{hyperref}
\usepackage{natbib}
\usepackage{graphicx}
\usepackage{ntheorem}
\usepackage{geometry}
\usepackage{caption}
\usepackage{indentfirst}

\newtheorem{defn}{Definition}

\nocite{*}

\title{Simulating evolution on fitness landscapes represented by Valued Constraint Satisfaction Problems}

\author{Alexandru Strimbu}

\date{Trinity term, 2019}

\begin{document}

\begin{titlepage}

\newcommand\blankpage{%
    \null
    \thispagestyle{empty}%
    \addtocounter{page}{-2}%
    \newpage}

%\pagecolor{white}%
  \LARGE
  \begin{center}
    {\bfseries \Huge{Simulating Evolution on Fitness Landscapes represented by Valued Constraint Satisfaction Problems}}\\[3em]
    \includegraphics[scale=0.5]{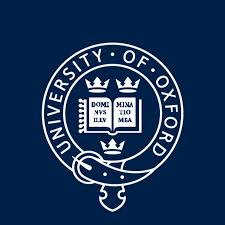}\\[3em]
    {\large
      \begin{tabular}{c}
      \\
      \\
      \\
      %\\
      \LARGE{\textbf{Alexandru Strimbu}}
      %\\
      \\
      \\
      \\
      \\
      \\
      \LARGE{St. Anne's College}
      \\
      \\
      \LARGE{University of Oxford}
      \\
      \\
      %\\
      \\
      \\
      %\\
      \\
      \\
      \LARGE{Trinity Term, 2019}
      \\
      \\
      %\\
      \\
      \\
      \\
      \end{tabular}%
    }
  \end{center}
\end{titlepage}

\begin{abstract}
Recent theoretical research proposes that computational complexity can be seen as an ultimate constraint that allows for open-ended biological evolution on finite static fitness landscapes. Whereas on easy fitness landscapes, evolution will quickly converge to a local fitness peaks, on hard fitness landscapes this computational constraints prevents evolution from reaching any local fitness peak in polynomial time.  Valued constraint satisfaction problems (VCSPs) can be used to represent both easy and hard fitness landscapes. Thus VCSPS can be seen as a natural way of linking the theory of evolution with notions of computer science to better understand the features that make landscapes hard. However, there are currently no simulators that study VCSP-structured fitness landscapes.

This report describes the design and build of an evolution simulator for VCSP-structured fitness landscapes. The platform is used for simulating various instances of easy and hard fitness landscapes. In particular, we look at evolution under more realistic assumptions than fittest mutant strong-selection weak mutation dynamics on the winding semismooth fitness landscape. The results obtained match with the theoretical expectations, while also providing new information about the limits of evolution. The last part of the report introduces a mathematical model for smooth fitness landscapes and uses it to better understand why these landscapes are easy.
\end{abstract}

\newpage

\section*{Acknowledgements}

I would like to thank my tutor and supervisor Peter Jeavons for reading the drafts of this project and offering valuable advice.
I would also like to express my sincerest gratitude towards my primary supervisor Artem Kaznatcheev, who provided invaluable support and expertise during our weekly meetings that ensured that the project goes forward.
Finally, I wish to thank my family for their continuous support and encouragement.

\newpage

\tableofcontents

\newpage

\section{Introduction}

\subsection{Motivation}

Ever since the $19^{\text{th}}$ century work of Charles Darwin, Alfred Russel Wallace and Gregor Mendel, biologists started doing experiments and building theories for finding the limits of biological evolution. An important question in this context is finding whether evolution can happen indefinitely (the so called "open-ended evolution"). Unfortunately, finding such answers can take a lot of resources and, most importantly, a lot of time, as it could take years to get enough generations of microorganisms (usually bacteria) for sound results. An example of this is Richard Lenski’s long-term E-coli evolution experiment (\cite{lenski}), which shows that evolution is still ongoing after $50000$ generations in the E-coli bacteria, even though the environment is static (\cite{wiser}). However, this research started in $1988$ and it is still in progress.

The advent of computers in the second half of the $20^{\text{th}}$ century provided useful tools for this field. The biologists could use powerful computational models to simulate the natural evolution and to get in a matter of hours enough data that would otherwise take years. Furthermore, theoretical computer science provided a series of theories that could be applied to evolution. This project covers both of these aspects, by building a simulator for finding experimental proof for a recent paper (\cite{artem2}) which claims that computational complexity is a constraint for biological evolution. 

Kaznatcheev's paper suggests that evolution can be seen as a computational problem, in which the organisms try to maximise their fitness over successive generations. This problem would still be constrained by the theory of computational complexity, which tells us that some problems are too hard to be solved in a reasonable amount of time. Unfortunately, this work is purely theoretical. My project comes in aiding with this, by actually building tools to simulate open-ended evolution on specific examples of both easy and hard fitness landscapes that arise from valued constraint satisfaction problems (VCSPs).

So, this project will touch three important aspects. Firstly, building an evolution simulator for VCSPs, considering the fact that currently there are no evolution simulators for VCSPs. Secondly, simulating both easy and hard problems and comparing the results with the theory and finally extending the theoretical work for some of the results found.

\subsection{Method and road-map}

The methods used to analyse VCSPs include an actual simulation of instances of VCSPs based on a stochastic evolutionary model, the computation of relevant statistics, and the comparison of some of the results with a deterministic model that can help make theoretical sense of our experiments. 

The simulator was built from scratch, as we needed a specific and easy to use platform for our VCSP experiments, that we could understand and alter if needed. A description of our simulator is presented in Section \ref{theSim}. The framework is then used to simulate evolution on specific instances of VCSPs (Section \ref{results}).

The results in Section \ref{results} are divided into results obtained for easy problems (the so-called "smooth landscapes"), in Section \ref{smoothLandscape}, and results for harder problems, in Section \ref{semiSmooth}. For the easy category, we present our simulation results, which agree with the theoretical results. For the harder category, we show the results and compare them with the easy landscapes.

Section \ref{model} introduces a new theoretical model for the smooth landscapes, that I developed for this project.

Finally, Section \ref{conclsuions} shows the conclusions of the project and suggests some further work that can be done on this topic.

The Appendices introduce a description of some mathematical topics that are used throughout the project report, together with a detailed description of the simulator's code.

\subsection{Contribution}

The main contribution of the project is an evolution simulator for VCSPs, which have not been previously studied in this context. The simulator is easy to use and configure by researchers in the field, opening the door for new possibilities for both biologists and computer scientists.

We show that we can use the simulator to encode experiments on smooth, semismooth and rugged fitness landscape (for definitions, see Section \ref{bioBack} and \cite{artem2}). We present a simulation of a candidate hard semismooth landscape (proved hard on fittest-mutant SSWM dynamics), which suggests that it might also be hard on more realistic evolutionary dynamics. We also show how the NK-model can be encoded in our simulator.

Finally, we introduce a mathematical model for smooth landscapes and compare it with the simulation results.

\newpage

\section{Background}

In this section we give standard definitions and results that will be used throughout the report. We start with the biology (Section \ref{bioBack}), then the computer science (Section \ref{vcspBackground}), as well as discussing current evolutionary simulators (Section \ref{avida}) and summarising recent theoretical work by \cite{artem2} (Section \ref{recRes}). Some similar examples appeared in one of my blog posts (\cite{strimbu}).

\subsection{Biology background} \label{bioBack}

This subsection presents some of the biological concepts that we will explore in this report. Many of the concepts used in biology do not have precise, mathematical definitions. As such, the presentation will be less formal than the mathematical and computational background.

The process of evolution is regarded as the transformation of the heritable traits of an organism, across successive generations (\cite{strickberger}). Those heritable characteristics are determined by the hereditary material. In this category, we will speak about the \textbf{genes}, which represent the basic information that dictates the features of an organism. The position of a gene inside the genetic code is known as a \textbf{locus}. Different variants of a gene can occupy the same locus, and these are called \textbf{alleles}. The total ensemble of genes forms the \textbf{genotype}. Biologists often think about genotypes as being implemented by the nucleic acids (DNA and RNA). In this context, an important concept is represented by \textbf{mutation}, which means a permanent alteration of the genome that can be passed on from one generation to the next. The space of possible mutations can be represented as an undirected graph with an edge from a genotype to its possible mutants, known as a \textbf{mutation graph}. Concrete biological definitions for these terms can be found in \cite{encyclopedia}.  
 
One of the key mechanisms of evolution is natural selection, which means that different organisms will survive and reproduce differently, due to different traits (\cite{strickberger}). This is closely related to the adaptability of the organisms to the environment. In this context, we can talk about \textbf{fitness} as a quantitative measure of natural selection, of the ability of the organism to reproduce and thrive in its environment (\cite{orr}). The \textbf{fitness function} will represent a mapping from a genotype to a numerical value which represents the fitness. Note that this should not be confused with the similarly named fitness functions of evolutionary game theory (\cite{rockne20192019}; \cite{kaznatcheev2017two}). Higher values will mean that the organism is a better fit for the environment. A \textbf{fitness landscape} is a form of visualising the relationship between the genotype and the fitness by combining the mutation graph and its fitness values (\cite{wright1932roles}). A value is assigned to each vertex, which represents a possible genome. We can see this graph as a "physical" landscape, with mountains, cliffs and valleys. A \textbf{local fitness peak} represents a peak in this landscape, a vertex in the graph for which all its neighbours do not have higher fitness value. A \textbf{fitness graph} can be seen as a representation of fitness landscapes as a directed graph, in which a vertex represents a possible genotype, and an edge is drawn from a lower fitness genotype to a mutationally adjacent higher fitness genotype(\cite{crona2013peaks}). 
 
 %%TC:ignore
\begin{figure}
\centering
\hspace*{-3.5cm}  
\begin{minipage}{\textwidth}
  \centering
  \includegraphics[scale = 0.85]{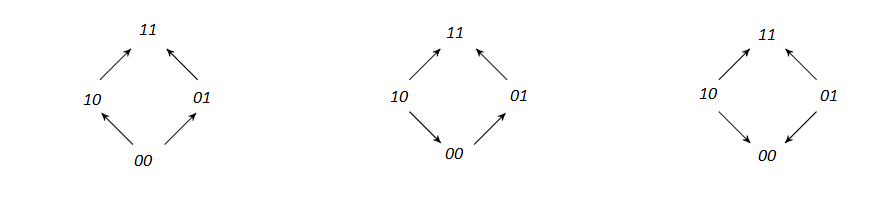}
\end{minipage}%
\caption{Three fitness graph representations of no epistasis (left), sign epistasis (center) and reciprocal sign epistasis (right). Edges are directed from lower to higher fitness}
\label{fig:fig20}
\end{figure}
%%TC:endignore
 
Another important notion is \textbf{epistasis}, which measures the interactions between loci in the fitness graph (\cite{weinreich2005perspective}). Depending of the type of interactions, we can have no epistasis, sign epistasis or reciprocal sign epistasis(\cite{weinreich2005perspective}, \cite{artem2}). consider two loci with alleles $0$ or $1$, with the combination $11$ being fitter than $00$. Then, we say that there is \textbf{no epistasis} between them if $f(11) > f(10) > f(00)$ and $f(11) > f(01) > f(00)$. \textbf{Sign epistasis} at the first locus happens when $f(11) > f(01)\ > f(00) > f(10)$. Finally, \textbf{reciprocal sign epistasis} between the 2 loci happens when $f(00) > f(10)$ and $f(00) > f(01)$ (Figure \ref{fig:fig20}).
 
We say that a landscape is easy, if we can get to the local optimal value in a reasonable (usually taken as polynomial) number of steps. Otherwise, we say that the landscape is hard (\cite{artem2}).
 
We will characterise the types of landscapes depending on the epistasis that they exhibit (\cite{artem2}). A landscape with no epistasis is known as a \textbf{smooth landscape} and has a single fitness peak. It is easy to navigate. A landscape with no reciprocal sign epistasis is known as a \textbf{semismooth landscape} and, while it also has a single peak, it can be hard to navigate. A landscape with reciprocal sign epistasis is known as a \textbf{rugged landscape}. It can be hard to navigate and it can also can have multiple peaks. 
 
The \textbf{NK model} (\cite{kauffman1989nk}) is a mathematical model for describing rugged fitness landscapes that are considered to be "tunable" by varying the two parameters $N$ and $K$, which change the size of the genome and the amount of epistasis. We define such a model as a fitness landscape on $\{0,1\}^{N}$. For a genome $x$, a gene locus $x_i$ is associated with $K$ other loci $x_1^i, ..., x_K^i$ such that we can define a positive fitness conribution $f_i(x_i, x_1^i, ..., x_K^i)$. Based on that, the total fitness is $f(x) = \sum_{i=1}^n f_i(x_i, x_1^i, ..., x_K^i)$.
 
The \textbf{selection coefficient} measures the difference in relative fitness between an organism and its possible mutants. The mathematical formula that we use for the selection coefficient is  
 
\[s = \frac{\overline{f^{N+}} - \overline{f}}{\overline{f}}\]
 
We define $\overline{f^{N+}}$ to be the average fitness of the "fitter" mutants (the mutants that are fitter than the organism, for each organism), and $\overline{f}$ to be the fitness average.
 
We will often speak about the \textbf{population}, which represents a collection of organisms that share the same environment. Thus, we are interested in how the entire population changes over time, as it navigates the fitness landscape. Under weak mutation, we can consider the population to be monomorphic (having a single genome), except for a short period of competition, before a new mutant becomes dominating. Furthermore, under strong selection, we can consider that any step in evolution will get us to a fitter mutant. Those rules are known as strong selection, weak mutation (\textbf{SSWM}) dynamics (\cite{orr2005genetic}, \cite{artem2}). So, we will try to ensure that, in the limit, we have a monomorphic population. As the population evolves over time, it will try to increase its fitness value, so it will navigate the fitness graph in an "uphill" fashion, by following the edges of the fitness graph. If the fittest possible mutant is always selected, we say that we have \textbf{fittest-mutant} dynamics, while if a mutant with higher fitness (but not necessarily the highest) is selected, we have \textbf{fitter-mutant} dynamics.
 
In this report we will describe systems where it is possible to keep getting higher and higher values for the fitness of our population throughout the simulation. We will consider such a system that does not reach a local fitness peak over a long time period to exhibit \textbf{open-ended evolution}.

\subsection{Computational background} \label{vcspBackground}

\begin{defn}[{{\cite{russell}, \cite{bernardo}}}]\ \\ A constraint satisfaction problem (CSP) is a mathematical question which asks for a set of objects whose state must satisfy certain given constraints. A CSP can be represented by $3$ sets: a finite set of variables $X$ (${X_1, ..., X_n}$), a finite set of domains $D$ that can be taken by the variables (${D_1, ..., D_n}$) and a finite set of constraints $C$. Each constraint ($C_i = (S_i, R_i)$) is represented by a set of variables ($S_i = (x_{i_1}, ... x_{i_m})$) that are taking part in the constraint (the \textit{scope}), and a relation that represents the values that can be taken by the variables in the scope ($R_i \subseteq D_{i_1} \times ... \times D_{i_m}$). A solution is represented by an assignment $v : X \mapsto D_1 \cup ... D_n$ that satisfies all the constraints $C_i$, such that $(v(x_i), ..., v(x_n)) \in R_i$.

\end{defn}

The number of variables ($m$) in the scope of each constraint represents the arity of that constraint. The arity of a CSP instance is the maximum arity over all its constraints. We can also represent binary constraints in the form of a graph with an edge between any two variables that interact. For constraints with higher arity, we will have hyper graphs where edges can join multiple variables.

A classic CSP is the problem of map colouring, in which the variables represent regions, the domain values are some colours, and the constraints define that the regions neighbouring each other should be coloured differently.

\begin{figure}
\centering
\hspace*{-1cm}  
\begin{minipage}{.5\textwidth}
  \centering
  \includegraphics[scale = 0.5]{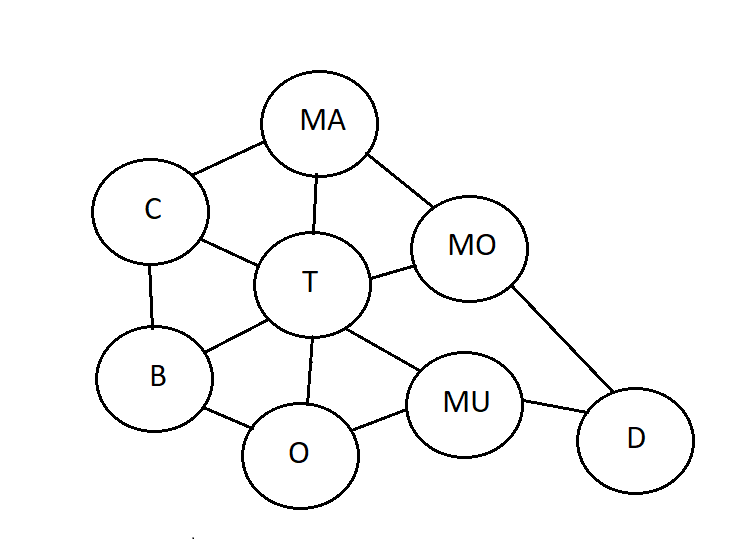}
\end{minipage}%
\begin{minipage}{.7\textwidth}
  \centering
  \includegraphics[scale = 0.5]{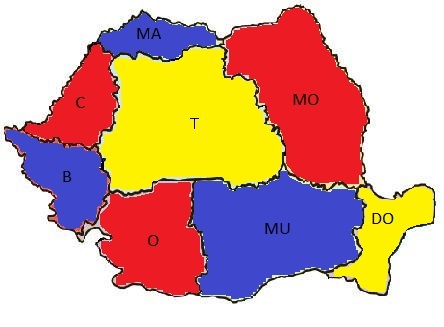}
\end{minipage}
\caption{The constraint graph (left) and the coloured map of Romania (right)}
\label{fig:fig1}
\end{figure}

A concrete example would be colouring the map of Romania. The variables are represented by Romania’s regions ($T$ – Transylvania, $MU$ – Muntenia, $MO$ – Moldavia, $D$ – Dobruja, $C$ – Crisana, $MA$ – Maramures, $B$ – Banat, $O$ – Oltenia) ($X = \{T, MU, MO, D, C, MA, B,O\}$), the domain is represented by some colours ($D = \{$ Red, Yellow, Blue $\}$), and the binary constraints represent $2$ neighbouring regions that should be coloured differently ($C = \{C \neq T, C \neq MA, C \neq B, MA \neq T, MA \neq MO, MO \neq T, MO \neq MU, MO \neq DO, DO \neq MU, MU \neq T, MU \neq O, O \neq T, O \neq B, B \neq T\}$). The constraint graph and a possible colouring are shown in Figure \ref{fig:fig1}.

%%TC:ignore
%%TC:endignore

We will now define a generalisation of the CSP.

\begin{defn}[{{\cite{jeavons}}}]\ \\ Let $f$ be a function $f : {D_1 \times ... D_m} \mapsto \mathbb{Z}$. A valued constraint $C$ over the set of variables $X$ is an expression of the form $f(x), x \in V ^ m$. The number $m$ represents the arity of the constraint. The sets $D_i$ represent domains.
\end{defn}

\begin{defn}[{{\cite{jeavons}}}]\ \\ A valued constraint problem (VCSP) is specified by a finite set $X$ of variables, a finite set $D$ of domains, a finite set of constraints $C$ and an objective function $\Phi$, such that $\Phi(x_1, ... x_n) = \sum_{i = 1}^{c} f_i(\mathbf{x_i})$, where each $f(\mathbf{x_i})$, with $1 \leq i \leq c$ is a valued constraint over $X$, as defined above, where $c$ is the number of constraints $C$. A solution is represented by an assignment that minimises the value of the function $\Phi$.
\end{defn}

If each constraint is represented by a function that returns a constant (the weight) when satisfied or $0$ otherwise, we obtain a subclass of VCSPs known as weighted constraint satisfaction problems (WCSPs).

An example of a WCSP instance would be a version of the map colouring problem presented above. If there are only two colours to work with, there is no proper colouring of the entire map. In other words, the CSP instance has no solution. Instead, suppose that we care more about some pairs of regions being coloured differently than others. In the case of the map of Romania, each constraint would also be given a weight, for example ($C =  \{(C \neq T, 2), (C \neq MA, 5), (C \neq B, 5), (MA \neq T, 1), (MA \neq MO, 5), (MO \neq T, 2), (MO \neq MU, 5), (MO \neq DO, 4), (DO \neq MU, 5), $ $(MU \neq T, 1), (MU \neq O, 5), (O \neq T, 2), (O \neq B, 5), (B \neq T, 1) \}$). Thus, we prefer adjacent regions to be coloured differently, but we care 5 times more about C and MA having different colours than we do about MA and T having different colours.

\begin{figure}
\centering
\begin{minipage}{.5\textwidth}
  \centering
  \includegraphics[scale = 0.5]{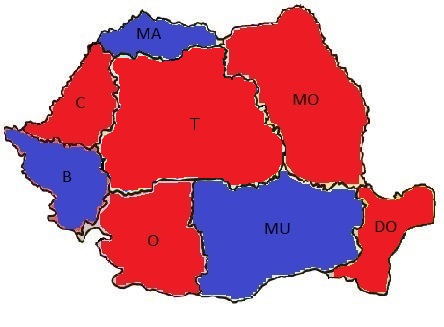}
\end{minipage}%
\begin{minipage}{.5\textwidth}
  \centering
  \includegraphics[scale = 0.5]{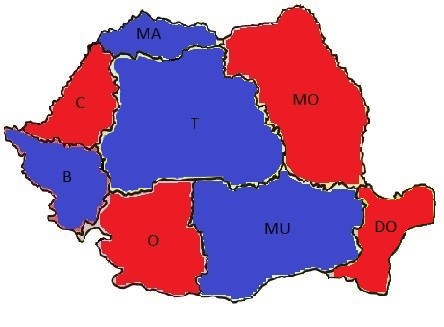}
\end{minipage}
\caption{Possible colourings of the map of Romania using only two colours}
\label{fig:fig2}
\end{figure}

Figure \ref{fig:fig2} left shows a colouring with a reward of $38$, while Figure \ref{fig:fig2} right has a total reward of $41$. 

%%TC:ignore
%%TC:endignore

An important complexity class for our discussion is the class \textbf{PLS}, which contains the problems for which locally optimal solutions can be verified in polynomial time (\cite{johnson1988easy}). As our evolutionary problems are local search problems, we are interested in knowing the difficulty of solving them, which requires finding a local optimal solution. It is generally accepted that PLS-complete problems (the hardest problems in PLS) can't be solved in polynomial time.

\begin{defn}[{{\cite{russell}, \cite{bernardo}}}]\ \\ The boolean satisfiability problem (SAT) is the problem of finding whether there exists a satisfiable assignment for a given boolean formula. 
\end{defn}

The weighted $2$-SAT problem is a variant of the $2$-SAT problem ($2$ variables per clause) that can be seen as a WCSP, with each clause representing a constraint that is satisfied if at least one variable is $true$ (\cite{artem2}). This problem is known to be PLS-complete (\cite{schaffer1991simple}).

We can see a VCSP as a genetic problem, in which $X$ is the set of loci, $D$ is the space of alleles, and $C$ is the interaction between genes (alleles at loci). The arity of a constraint is analogous with the degree of epistasis in biology. Binary (i.e. arity $2$) constraints are like pairwise epistasis, and higher arity constraints are like higher-order epistasis. The constraint graph can be seen as a gene interaction network (\cite{artem1}). Furthermore, in a biological setting, we think of the total reward associated with a variable assignment as the fitness of the corresponding genotype (Table \ref{table:1}).
See \cite{artemCP} for a more detailed discussion of the mathematics of representing fitness landscapes as VCSPs.

%%TC:ignore
\begin{table}[]
\centering
\begin{tabular}{ |c|c| }
 \hline
 \textbf{Computational (VCSP) terms} & \textbf{Evolutionary terms} \\ 
 \hline
 Variable & Locus \\  
 \hline
 Single variable assignment & Gene \\
 \hline
 All variable assignment & Genotype \\
 \hline
 Constraints & Gene interactions \\
 \hline 
 Constraint weight & Gene interaction’s fitness contribution \\
 \hline
 Constraint graph & Gene interaction network \\
 \hline
 Arity & Degree of epistasis \\
 \hline
 Total reward & Fitness\\
 \hline
\end{tabular}
\caption{Terminology translation between Computer Science and Biology}
\label{table:1}
\end{table}
%%TC:endignore

\subsection{The Avida artificial life simulator} \label{avida}

While there are currently no other biological evolution simulators for VCSP-structured fitness landscapes, there exist some more general artificial life programs for studying evolutionary biology. An example of this is the Avida platform (\cite{ofria2004avida}), which is widely used by biologists for their computational experiments. A recent paper (\cite{wiser2018boundedness}) uses Avida to argue that the simulated environment can exhibit open-ended dynamics.

However, this type of software is too general for our purposes. Trying to add VCSPs to this platform, together with all the other existing functions of the system will make the simulator very inefficient, and larger experiments will be unfeasible. In the same time, the system will be too hard to analyse rigorously. So, it was better to completely engineer a new, simpler program than trying to force an existing system to be able to solve VCSPs, while carrying a huge amount of overhead that was not needed.

\subsection{Recent theoretical results on open-ended evolution} \label{recRes}

One reason we focus on VCSPs is because they can express the reduction that \cite{artem1} used when proving that the NK-model is PLS-complete. He focused on the weighted 2-SAT problem, and showed how it can be encoded within the NK-model. Thus, he showed that a weighted 2-SAT instance with variables $x_1,…, x_n$, clauses $C_1, …,C_m$ and positive integer costs $c_1, … c_m$ can be encoded in the NK-model by building a landscape with $m + n$ loci, with the first $m$ labelled $b_1, …, b_m$ and the next $n$ labelled $x_1, …, x_n$. \cite{artem1, artem2} defined the corresponding fitness effect of the locus as:

%%TC:ignore
\begin{quote}
\begin{center}
\begin{tabular}{ l }
    $f_k(0x_ix_j) = \left\{
                        \begin{array}{ll}
                          c_k \textrm{\quad if } C_k \textrm { is satisfied} \\
                          0 \textrm{\quad otherwise}\
                        \end{array}
                      \right.$ \\
                      
    $f_k(1x_ix_j) = f_k(0x_ix_j) + 1 $
\end{tabular}
\end{center}
\end{quote}
%%TC:endignore

Since it can express weighted 2-SAT, the NK-model is hard for all evolutionary dynamics.

Another important result that we will use in Section \ref{semiSmooth} appears in \cite{artem2}, and provides a way of recursively constructing a hard semismooth landscape under fittest-mutant SSWM dynamics:

Assuming that we have a semismooth fitness landscape on $\{0,1\} ^ m$ with fitness function $f$ which takes $k$ steps to reach an evolutionary equilibrium $x^*$, starting from $0^m$, we can construct a semismooth fitness landscape on $\{0,1\} ^ {m + 2}$ with fitness function $f'$, which takes $2(k + 1)$ steps to reach its equilibrium by starting from $0^{m+2}$:

\[f'(xab) = \left\{
                \begin{array}{ll}
                    f(x) \textrm{\hspace{2.8cm} if } a = b = 0 \\
                    f(x) + s^- \textrm{\hspace{2cm} if } a \neq b \textrm{ and } x \neq x^* \\
                    f(x^*) + s^- \textrm{\hspace{1.85cm} if } a = 0, b = 1 \textrm{ and } x = x^* \\
                    f(x^*) + s^+ \textrm{\hspace{1.85cm} if } a = 1, b = 0 \textrm{ and } x = x^* \\
                    f(x \oplus x^*) + f(x^*) + 2s^+ \textrm{\quad if } a = b = 1\\
                \end{array}
            \right.\]
          
Where we have:

\[s^+(x) = \textrm{max}_{y \in N(x), f(y)>f(x)} f(y) - f(x)\]
\[s^-(x) = \textrm{min}_{y \in N(x), f(x) + s^+(x)>f(y)>f(x)} f(y) - f(x)\]
\[s^+ = \textrm{min}_{x} s^+(x)\]
\[s^- = \textrm{min}_{x} s^-(x)\]
\[N(x) = \textrm{neighbours of x}\]
(as defined in \cite{artem2}, Appendix E)

The initial function can be taken to be $f(00) = 2, f(01) = 3, f(10) = 4, f(11) = 6$. We can use the above recursion to build a semismooth landscape with $2n$ loci for which we need $2^{n+1}-2$ fittest mutant steps to reach the peak at $0^{2n-2}11$ by starting from $0^{2n}$.

\newpage

\section{The evolution simulator} \label{theSim}

This section introduces a general description of the simulator and its simulation methodology. The detailed description and some important parts of the code can be found in Appendix D (Section \ref{appD}). The complete simulator can be found on GitHub at \url{https://github.com/strimbu-alexandru/Evolution-Simulator} (\cite{alexCode}).

\subsection{Description of the simulation methodology}

The simulation is done by starting with an initial configuration, which includes the initial population. For each organism in the population, we compute the value of the fitness function (which is fixed to be greater than or equal to 1, and represents the expected number of offspring). The simulation proceeds round by round, by generating the next population from the current one. This is done in three steps. Firstly, we sample, for each organism in the current population, a number using a Poisson distribution, parametrised by the value of the fitness function, giving us the number of offspring. Secondly, as we keep the population size constant, we sample from the hypergeometric distribution (Appendix B), getting the number of children that survive to the next generation for each organism. Finally, we replace the population with the new organisms that are obtained by copying the genome of their parents. During this stage, each genome can suffer mutations with a given probability rate. We implemented two mutation methodologies. The default one, which is used in most of our experiments, decides if a mutation happens or not for the entire genome, and if the answer is yes, it randomly chooses a gene (constraint variable) that will change to another allele (domain value). This ensures that at most a single gene can mutate in a single step. The second possibility is to do this process for each gene, deciding if it mutates or not. This can give rise to many mutations in one step.

The above process finishes after a given number of rounds. For each round, we compute a set of statistics, which are plotted on graphs and will be used in  the next section to present the results of our experiments. A description of some of the statistics concepts used can be found in Appendix A.

\subsection{The design of the simulator}

The evolution simulator is written in Python (\cite{python}) and designed for solving VCSPs using an evolutionary method and collecting and presenting statistics about the organisms. The code was prototyped and the experiments were ran using the Jupyter Notebook (\cite{jupyter}), which helped adopting an Agile approach. For many of our computations, we use the package NumPy (\cite{numpy}). I decided to use this programming language because it is widely used by both computer scientists and non-computer scientists (like biologists that might be interested in this simulation software), there is a large and active community around it, and there are many tools for visualising the results (through graphs) and for computing statistics. 

\begin{figure}
\begin{minipage}{.5\textwidth}
  \includegraphics[scale = 0.47]{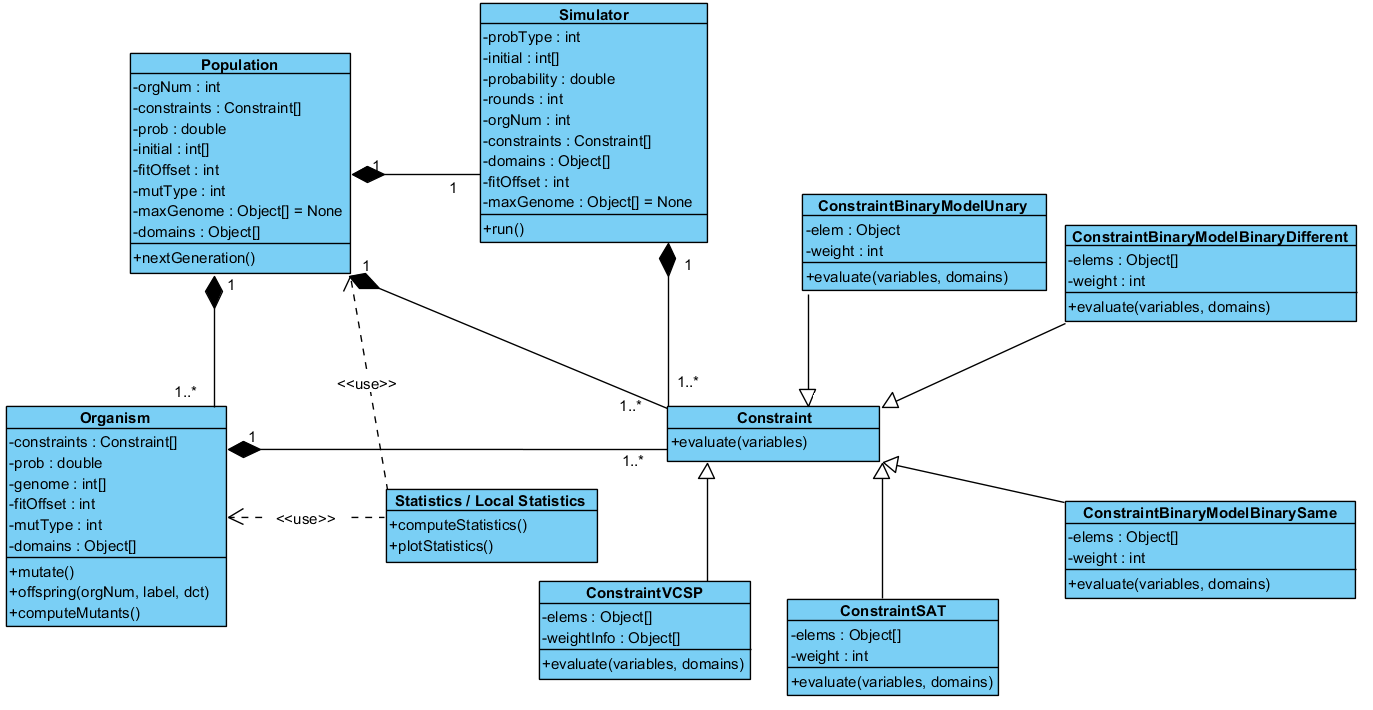}
\end{minipage}%
\caption{An UML diagram specifying the design of the simulator}
\label{fig:fig22}
\end{figure}

We built the code using  standard conventions of object-oriented programming. This makes it easy to understand and customise. Following this paradigm, the program is structured into eleven classes: \textit{Simulator}, \textit{Population}, \textit{Organism}, \textit{Constraint} (and the derived classes \textit{ConstraintSat}, \textit{ConstraintBinaryModelUnary}, \textit{ConstraintBinaryModelBinaryDifferent}, \textit{ConstraintBinaryModelBinarySame}, \textit{ConstraintVCSP}),  \textit{Statistics} and \textit{LocalStatistics}. Detailed descriptions and parts of the code can be found in Appendix D. The UML diagram of our simulator can be seen in Figure \ref{fig:fig22}. 

%%TC:ignore
%%TC:endignore

\subsubsection{The \textit{Simulator} class}

The \textit{Simulator} class (detailed in Section \ref{simulator}) represents the interface between the user and the back end of our simulator. It is responsible for obtaining all the user-defined inputs and configurations. Calling the \textit{run} method of this class will begin the simulation round by round by producing the next generation through a call of the \textit{Population} class. In this way, we ensure a good level of abstraction that makes our code easy to understand and modify. Finally, this class returns the output of the simulator in the form of statistics computed by the two components responsible for that.

\subsubsection{The \textit{Population} class} \label{secPopul}

The \textit{Population} class (detailed in Section \ref{population}) represents the abstraction for the total population of organisms. Its most important function is computing the next generation based on the current one. For this, we obtain the pool of possible offspring for each organism (which is not empty), and then select from them for the next round a number equal to the size of the population (as our population should remain the same).

For each organism, we get the number of offspring, and use the multivariate hypergeometric distribution to find the number of surviving children for each individual. As we could not find a package that considered the multivariate case, we had to implement it by repeatedly sampling from the bivariate distribution. Finally, we only mutate the surviving children.

\subsubsection{The \textit{Organism} class}

The \textit{Organism} class (detailed in Section \ref{organism}) represents the abstraction of an organism inside the population. It is responsible for a number of different operations.

Firstly, it calculates the fitness value of the organism, by calling the corresponding evaluation function in each constraint.

Secondly, it computes a mutation of the organism's genome. Depending on the user's input, it selects from the two mutation possibilities described at the beginning of this chapter. We sample from a Bernoulli distribution with the mutation probability as parameter to decide if a mutation happens. Then, we choose randomly a new domain value to mutate to.

Thirdly, this class returns the number of children for the next generation. For this, we sample from a Poisson distribution, having as a parameter the fitness of the organism (with a small correction to ensure that we have at least one child). We use this distribution as the expected value is the same as the parameter given.

Finally, there is a method for returning the possible mutants that have their genome at a Hamming distance of $1$ from the organism's genome.

\subsubsection{The \textit{Constraint} class and its subclasses}

The \textit{Constraints} class (detailed in Section \ref{constraints}) is an abstract class which represents the blueprint of encoding a constraint. This structure permits an easy and consistent way of handling various types of constraints, and makes them easily customisable. In principle, the heart of a constraint is the evaluation function, which returns a value based on the variables given. For our experiments we use three constraint types (the general valued CSP, the binary / unary constraints and the SAT constraints).

\subsubsection{The statistics classes}

Finally, we present the two statistics classes that are responsible for providing the output of our simulator.  We use the Matplotlib (\cite{matplotlib}) package to visualise those statistics. 

The \textit{Statistics} class (detailed in Section \ref{genStats}) is responsible for the general fitness statistics of the population (like the average fitness and some variations). Those are computed and stored after each round, and plotted at the end of the simulation. 

The \textit{LocalStatistics} class (detailed in Section \ref{localStatsClass}) is responsible for some of the "local" statistics, which depend on the organisms and their neighbouring mutants (like fitness and selection coefficient). After each round of simulation, statistics are computed and stored in a dictionary. This is then used for getting the results.

The local fitness statistics are obtained similarly to the general ones. However, the computation of the selection coefficient involves calculating the actual values and fitting an exponential function to the data, which is used in interpreting the results. For this, we use the Scipy package (\cite{scipy}), with a fitting method that minimises the least square function. For a better fit, we compute the standard error (using the propagation of uncertainty, described in Appendix A) and use this to weight the points differently according to their error. Moreover, we show log-plots of the selection coefficient for better analysis purposes. 

Finally, we compute the (minimum and average) distance to the peak. This is done by calculating the Hamming distance between the genome of the organisms and the given peak.

\subsection{Challenges and limitations} \label{challenge}

Probably the most important challenge was completely designing and building a VCSP simulator, because there are currently no simulators for general VCSP - structured fitness landscapes, as they were not studied before in this context.

An obvious challenge encountered when building such a piece of software is trying to keep a balance between the efficiency of the software, which is crucial for running bigger and more meaningful experiments, the statistics that are computed, which affect the efficiency, but are needed for a better understanding of the results, and finally the general usability of the software, together with the possibility of customisation.

In order to balance all these aspects, we adopted a modular design that is easy to understand and makes the customisation of some specific components easier. We implemented some default statistics, with a few of them that can be ignored to make the program more efficient and the possibility of adding more by the user, if required. Finally, in order to make the software as efficient as possible, we used some optimisation techniques like delaying some of the computations until when they are needed and storing some of the data in dictionaries or lists to avoid unnecessary computations. 

An example of this kind of optimisation can be seen when computing the next generation of organism. The first strategy considered (that also mimics nature) was getting all the offspring (that already suffered possible mutations) and selecting a fixed number out of them in a random way to survive for the next round. However, this approach was very inefficient, as it wastes a lot of time and memory that is just discarded at the end. Our second strategy, that is described in Section \ref{secPopul}, avoids this problem by using the hypergeometric distribution. This technique reduced the time needed for computing bigger experiments from days to hours.

\begin{figure}
\begin{minipage}{\textwidth}
    \centering
    \includegraphics[scale = 0.7]{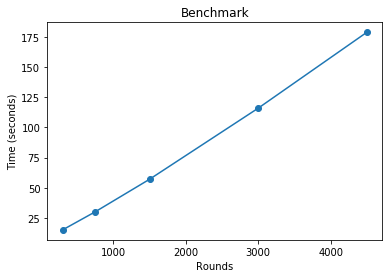}
\end{minipage}
\caption{Time taken to finish simulation when increasing the number of rounds for $\mu = 0.01$, $orgNum = 200$ and $genomeLen = 20$. Observe that the increase is linear.}
\label{fig:fig23}
\end{figure}

However, we admit that the system has certain limitations regarding the size of the experiments that can be performed (Figure \ref{fig:fig23}). As this simulator is the first one to investigate this area, we expect improvements to be possible.

%%TC:ignore
%%TC:endignore

Some changes that appear to be possible include exploiting computational parallelism by computing some organism specific data in parallel or rewriting the program in a more efficient framework (losing some of the benefits, such as usability or customisation).

\newpage

\section{Experimental Results} \label{results}

\subsection{Smooth landscapes} \label{smoothLandscape}

We model a smooth landscape as a VSCP starting with an initial genotype of $0$s and reaching a genotype of $1$s. The fitness value of an organism is represented by the number of $1$s in its genome. We use the setting in which a mutation can change at most a single gene, so every mutation can increase or decrease the fitness by $1$. Thus, the constraints are represented as unary SAT constraints, the setting being identical with a weighted satisfiability problem that has a single clause for each variable.

%%TC:ignore
\begin{minted}{python}
initial = [0] * 8
clause1 = ConstraintSat([1],1)
clause2 = ConstraintSat([2],1)
clause3 = ConstraintSat([3],1)
clause4 = ConstraintSat([4],1)
clause5 = ConstraintSat([5],1)
clause6 = ConstraintSat([6],1)
clause7 = ConstraintSat([7],1)
clause8 = ConstraintSat([8],1)
clauses = [clause1, clause2, clause3, clause4, clause5, clause6,
    clause7, clause8]
\end{minted}
%%TC:endignore

We formulate such a problem in our simulator, for a mutation probability of $0.01$, $200$ organisms and a run of $300$ rounds. Using a small probability will minimise the error caused by many mutations (so we can consider the experiment as having weak mutation, which implies a monomorphic population). This ensures that the average (normalised) fitness gets close to $1$, and the selection coefficient close to $0$. We present the results for a typical run of the simulator.

%%TC:ignore
\begin{figure}
\begin{minipage}{.5\textwidth}
  \includegraphics[scale = 0.7]{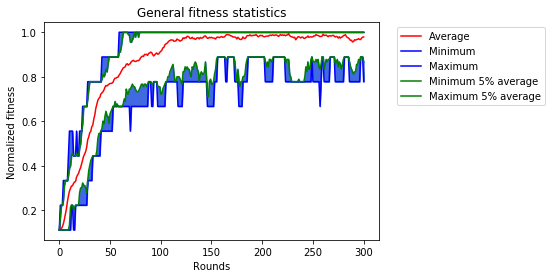}
\end{minipage}
\caption{The fitness statistics for a smooth landscape with $\mu = 0.01$, $orgNum = 200$, $rounds = 300$ and $genomeLen = 8$}
\label{fig:fig4}
\end{figure}
%%TC:endignore

%%TC:ignore
\begin{minted}{python}
probability = 0.01
rounds = 300
orgNum = 200
mySim = Simulator(1, initial, probability, rounds,
    orgNum, clauses, None, 1, 1, True)
mySim.run()
mySim.printStatistics()
mySim.printLocalStatistics()
\end{minted}
%%TC:endignore

The general fitness statistics (Figure \ref{fig:fig4}) show that our population converges fast to its local optimum. The first organism reaches optimality after about $60$ rounds, and the entire population reaches an average fitness close to $1$ after about $100$ rounds, after which the population stabilises.

%%TC:ignore
\begin{figure}
\begin{minipage}{.5\textwidth}
  \includegraphics[scale = 0.60]{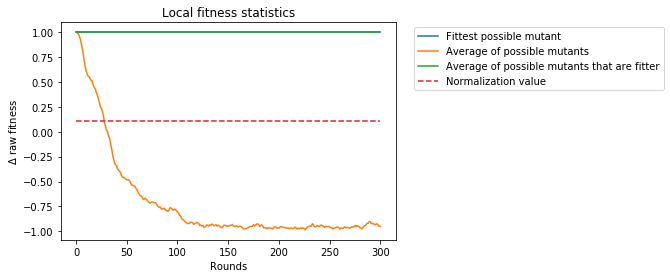}
\end{minipage}
\caption{The local fitness statistics for a smooth landscape with $\mu = 0.01$, $orgNum = 200$, $rounds = 300$ and $genomeLen = 8$}
\label{fig:fig5}
\end{figure}
%%TC:endignore

The local fitness statistics (Figure \ref{fig:fig5}, plotted against a "delta raw fitness" (the difference in fitness between an organism and a possible mutant; 'raw' meaning not normalised), shows, as expected, that, at every step, we have mutants that are fitter, and the difference is $1$. It can be observed that the average delta fitness of the possible mutants decreases over time. As the population gets closer to the optimal value and it can't increase its fitness anymore, the mutations can only bring a decrease in fitness. However, the population remains very close to the optimum, as the natural selection, implemented by the offspring generation, favours the fitter organisms.

%%TC:ignore
\begin{figure}
\begin{minipage}{.5\textwidth}
  \includegraphics[scale = 0.35]{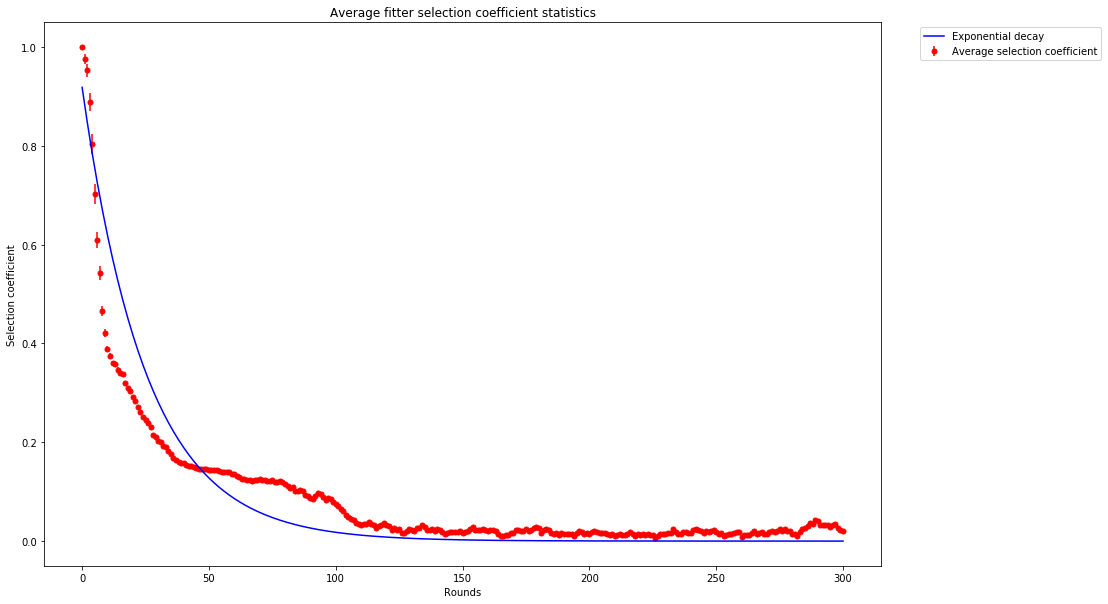}
\end{minipage}
\caption{The selection coefficient for a smooth landscape (red) and a best exponential fit (blue), simulated for $\mu = 0.01$, $orgNum = 200$, $rounds = 300$ and $genomeLen = 8$}
\label{fig:fig6}
\end{figure}
%%TC:endignore

\begin{figure}
\begin{minipage}{.5\textwidth}
  \includegraphics[scale = 0.65]{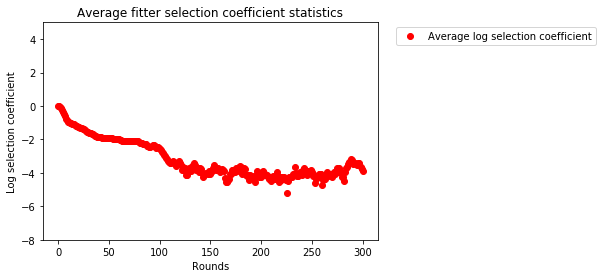}
\end{minipage}
\caption{The selection coefficient for a smooth landscape, plotted on a log-y scale, with $\mu = 0.01$, $orgNum = 200$, $rounds = 300$ and $genomeLen = 8$}
\label{fig:fig7}
\end{figure}

The average selection coefficient plot (Figure \ref{fig:fig6}) and the log-plot (Figure \ref{fig:fig7}) confirm the expected theoretical results (\cite{fisher1930genetical}, \cite{artem2}), that the selection coefficient should see an exponential decay for smooth landscapes. We observe that the coefficient starts from a value of $1$ and decreases, having a decay that can be approximated by an exponential function, reaching a value close to $0$ (this corresponds with the average fitness from the previous graph getting close to $1$).

%%TC:ignore
%%TC:endignore

\begin{figure}
\begin{minipage}{.5\textwidth}
  \includegraphics[scale = 0.65]{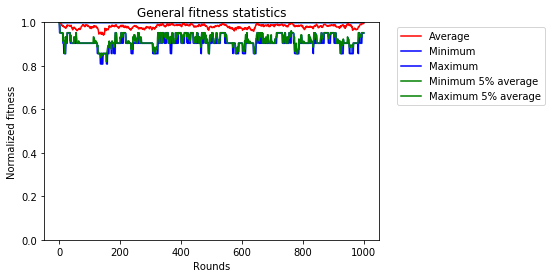}
\end{minipage}
\caption{A smooth landscape simulation starting from peak with $\mu = 0.01$, $orgNum = 200$, $rounds = 300$ and $genomeLen = 8$}
\label{fig:fig12}
\end{figure}

Finally, we show that the optimal value is stable, by starting the simulation at the peak and observing that it remains around the maximum value of $1$ (Figure \ref{fig:fig12}).

%%TC:ignore
%%TC:endignore

\subsection{Semismooth landscapes} \label{semiSmooth}

Next, we construct a hard semismooth fitness landscape, as described in Section \ref{recRes}. This is done by implementing the fitness function recursively and encoding it in our simulator as a single $n-ary$ constraint (denoted as $myFunc$; the code for it can be found in Appendix E). 
It is unlikely that this could be implemented by significantly simpler constraints, as argued in \cite{CCKW19}.
The simulation is done with a small probability of mutation of $0.01$ for a high number of $3000$ rounds, with a genome of length $20$. We start with a genome made only of $0$s, while the maximum fitness value is attained at a configuration of $18$ $0$s and $2$ $1$s. 

%%TC:ignore
\begin{minted}{python}
initial = [0] * 20
probability = 0.01
rounds = 1000
orgNum = 100
maxVal = 1022000003072  #maximum value that can be attained
cons = ConstraintVCSP([0,1,2,3,4,5,6,7,8,9,10,11,12,13,14,15,16,
    17,18,19],(myFunc,maxVal))
constraints = [cons]

mySim = Simulator(1, initial, probability, rounds,
    orgNum, constraints, None, 1, 1, True)
mySim.run()
mySim.printStatistics()
mySim.printLocalStatistics()
\end{minted}
%%TC:endignore

%%TC:ignore
\begin{figure}
\centering
\begin{minipage}{1\textwidth}
  \centering
  \includegraphics[scale = 0.7]{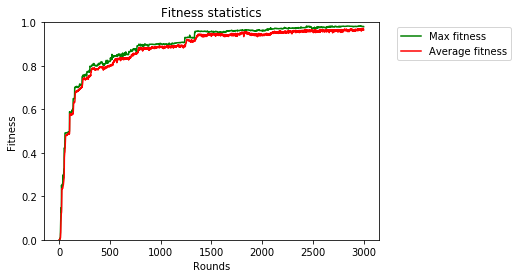}
\end{minipage}
\begin{minipage}{1\textwidth}
  \centering
  \includegraphics[scale = 0.7]{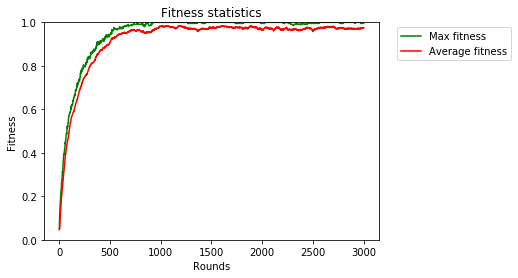}
\end{minipage}
\caption{Plots of the average (normalised) fitness for each round, for the semismooth landscape (top) and the smooth landscape (bottom), both with $\mu = 0.01$, $orgNum = 200$, $rounds = 3000$, $genomeLen = 20$}
\label{fig:fig10}
\end{figure}
%%TC:endignore

We plot the average fitness and maximum fitness at each round for $10$ different runs that are then averaged (Figure \ref{fig:fig10} (top)).

We observe that, although we require only $2$ mutations (of the last $2$ genes) to reach the optimum, our population does not reach it (for all $10$ runs that are averaged) even after $3000$ rounds. Actually, it requires $300$ rounds to reach $60\%$ of the max fitness and almost $600$ rounds to reach $80\%$ of it. After $1000$ rounds, it gets close to $90\%$, and it stays there until we reach $1500$ rounds. After that, it stays between $95\%$ and $100\%$, without actually reaching it, getting very close after $2000-2500$ rounds.

We can compare the above results with a similar simulation of a smooth fitness landscape similar to the one defined in Section $4.1.1$, but having a genome length of $20$ (Figure \ref{fig:fig10} bottom). In this case, it arrives at $60\%$ of the max fitness after $150$ rounds, at $80\%$ after $300$ rounds and it reaches the maximum fitness at about $750$ rounds (for all $10$ runs that are averaged), and after that it stays around the optimum. So, while for the smooth landscape the simulation arrives fast at the optimum value, this does not happen in the semismooth case. We mention that our simulation does not implement a strict fittest mutant SSWM dynamics, so the result we obtained is a surprising and important one, as our simulation methodology is closer to how nature works. By using a small mutation probability, we can consider that strong selection happens, as we reduce the not-so-fit organisms to a minimum. From the way in which we select our next generation based on fitness, we can say that we use a fitter mutant approach.

%%TC:ignore
\begin{figure}
\begin{minipage}{.5\textwidth}
  \includegraphics[scale = 0.7]{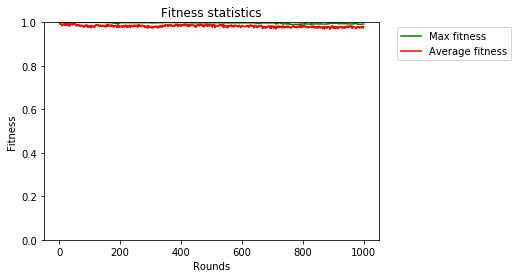}
\end{minipage}
\caption{Plot of the fitness statistics for the semismooth landscape starting at the peak, with $\mu = 0.01$, $orgNum = 200$, $rounds = 3000$, $genomeLen = 20$}
\label{fig:fig13}
\end{figure}
%%TC:endignore

As we have shown for the smooth landscape, we can note that the semismooth landscape is also stable at the peak. Thus, if we start the simulation at the optimum level, we remain around a fitness of $1$ for the entire simulation (Figure \ref{fig:fig13}).

%%TC:ignore
\begin{figure}
\centering
\begin{minipage}{1\textwidth}
  \centering
  \includegraphics[scale = 0.65]{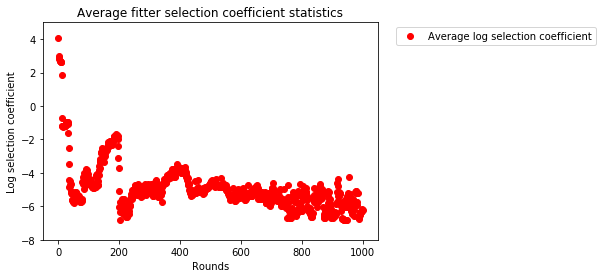}
\end{minipage}
\begin{minipage}{1\textwidth}
  \centering
  \includegraphics[scale = 0.65]{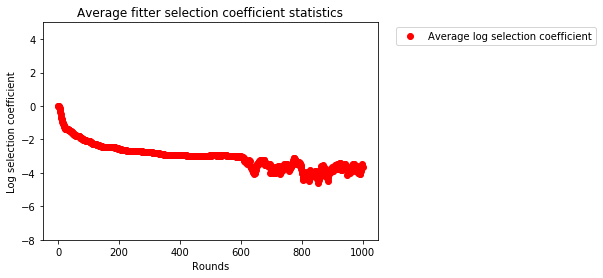}
\end{minipage}
\caption{Plots of the log-selection coefficients for each round, for the semismooth landscape (top) and the smooth landscape (bottom), with $\mu = 0.01$, $orgNum = 200$, $rounds = 3000$, $genomeLen = 20$}
\label{fig:fig11}
\end{figure}
%%TC:endignore

A further comparison can be made between the selection coefficients of the smooth and semismooth landscapes. We can look at the graphs of log-selection coefficient for the two types of landscapes on a typical run of the simulator for $1000$ rounds (Figure \ref{fig:fig11}). The graph for the semismooth landscape (top) shows some "bumps", explained by the sudden increases in fitness that happen when a better mutant is found. This happens after the fitness stabilises, so the selection coefficient gets smaller. The graph for the smooth landscape (bottom) follows the approximate path of a line (on the log-plot), confirming an exponential decay and the fact that we have a steady increase in the fitness of the population.

%%TC:ignore
\begin{figure}
\centering
\begin{minipage}{1\textwidth}
  \centering
  \includegraphics[scale = 0.67]{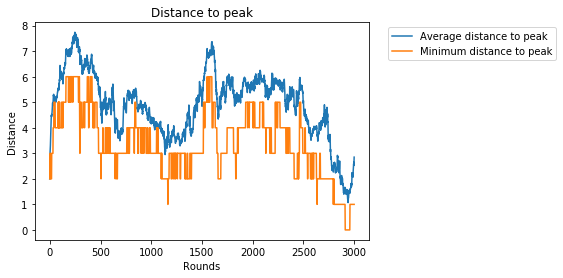}
\end{minipage}
\begin{minipage}{1\textwidth}
  \centering
  \includegraphics[scale = 0.67]{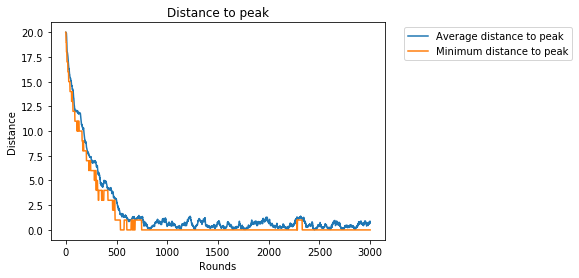}
\end{minipage}
\caption{Plots of the average and minimum distance to the peak for each round, for the semismooth landscape (top) and the smooth landscape (bottom), with $\mu = 0.01$, $orgNum = 200$, $rounds = 3000$, $genomeLen = 20$}
\label{fig:fig14}
\end{figure}
%%TC:endignore

Finally, we can compare the average and minimum distance to the (known) peak for both landscapes on typical runs (Figure \ref{fig:fig14}). This shows a crucial difference between the two runs. Thus, from the top graph, showing the run on the semismooth landscape, we observe that, even if the fitness of the population increases, we actually get to a population that is very different from the peak. Starting from organisms that are different from the optimal genome only at $2$ genes, we get to an average distance to the peak of $8$ genes. So, navigating the landscape uphill, the population moves along a very long path towards the peak, that was initially at a distance of just $2$ genes. After about $2900$ rounds, the simulation reaches the peak (i.e. the distance to the peak for some organism is $0$).

On the other side, looking at the lower figure, we observe that on the smooth landscape, the distance to the peak has a steady decrease, similar to the steady increase in fitness. This shows that the population navigates to the peak by getting closer and closer to it, on a short path.

Using the formula from Section \ref{recRes} with $n=10$, we get that there are $2^{10+1} - 2 = 2046$ fittest mutant steps for SSWM dynamics on this semismooth landscapes. So our results for getting to or close to the peak after 2000 rounds of simulations are consistent with this.

\subsection{Rugged landscapes}

In this section, we present some results based on rugged landscapes and a possible encoding of the NK-model in our simulator.

\subsubsection{An easy rugged landscape}

A rugged landscape can be generated in the simulator using the binary constraints formulation. This is done by generating unary constraints for each gene and binary constraints for each pair of genes. Using a genome length of $8$, we will need $8$ unary constraints and $28$ binary constraints. We first randomly generate $8$ weights between $0$ and $50$ for the unary constraints and $28$ weights between $-50$ and $50$ for the binary constraints.

%%TC:ignore
\begin{minted}{python}
weights = []
for i in range(8):
    weights.append(randint(0, 50))
for i in range(28):
    weights.append(randint(-50, 50))
\end{minted}
%%TC:endignore

For our example, we obtained 

%%TC:ignore
\begin{minted}{python}
weights = [31, 13, 16, 36, 32, 45, 22, 49, -32,
    -50, 49, -48, -18, -10, -20, 31, 14, -26, 43,
     8, 34, 1, 35, -16, -1, -9, -7, -35, -7, -44,
     -2, -1, 43, -6, 50, -6]
\end{minted}
%%TC:endignore

Afterwards, we define the simulation setting. For the binary constraints, if the weight is negative, it will be applied (in absolute value) as a "different" constraint, meaning that if the $2$ loci have different values, the weight is applied. Otherwise, it is applied as a "same" constraint, but only if both genes are $1$. We start the simulation from the fitness minimum of all $0s$.

%%TC:ignore
\begin{minted}{python}
initial = [0] * 8
probability = 0.01
rounds = 500
orgNum = 100

constraints = []
l = 0

for i in range(8):
    constraints.append(ConstraintBinaryModelUnary(i, weights[l]))
    l += 1
    
for i in range(7):
    for j in range(i + 1, 8):
        w = weights[l]
        l += 1
        if w < 0:
            constraints.append(ConstraintBinaryModelBinaryDifferent(
                [i,j],[-w,-w]))
        else:
            constraints.append(ConstraintBinaryModelBinarySame(
                [i,j],[0,w]))

mySim = Simulator(2, initial, probability, rounds, orgNum,
    constraints, None, 1, 1, False, [0, 1, 1, 0, 1, 1, 1, 1])
mySim.run()
mySim.printStatistics()
mySim.printLocalStatistics()
\end{minted}
%%TC:endignore

In this setting, it can be observed that between the first $2$ loci exists reciprocal sign epistasis (see Section \ref{bioBack}). The fitness values for the four possible assignments are $f(00) = 0$, $f(10) = 31 + 32 = 63$ (from the binary constraint and one unary constraint), $f(01) = 32 + 13 = 45$, $f(11) = 31 + 13 = 44$ (from both unary constraints).

%%TC:ignore
\begin{figure}
\centering
\begin{minipage}{1\textwidth}
  \centering
  \includegraphics[scale = 0.68]{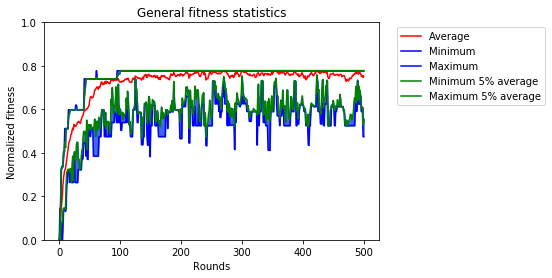}
\end{minipage}
\begin{minipage}{1\textwidth}
  \centering
  \includegraphics[scale = 0.68]{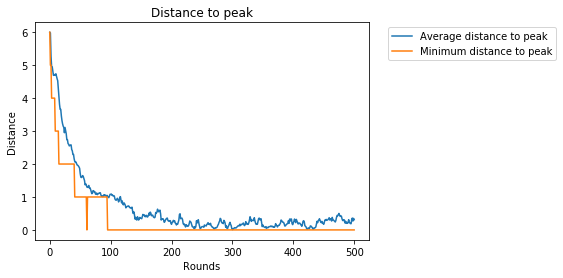}
\end{minipage}
\caption{Plots of the fitness data (top) and the distance to the peak (bottom) for the rugged landscape }
\label{fig:fig19}
\end{figure}
%%TC:endignore

However, the results that we obtain Figure (\ref{fig:fig19}) show us that in practice this is an easy landscape. Thus, the population reaches its maximum attainable fitness (which is about 80\% of the maximum weight for all constraints) fast, after about $150$ rounds. This is also shown by the second graph, which represents the distance to the peak. So, we conclude that this landscape, even though it is a rugged landscape, is still easy to navigate.

\subsubsection{Simulating the NK model}

The NK model defined in Section \ref{bioBack} can be represented in our simulator using VCSP constraints. Thus, for the locus $x_i$ and the $K$ linked loci $x_1^i, ..., x_K^i$, the fitness contribution $f_i = (x_i, x_1^i, ..., x_K^i)$ can be written as a VCSP constraint, defined for the corresponding positions in the genome and the corresponding function. So, as the fitness equals the sum of the contribution of each constraint, the final fitness in our simulation corresponds with the one of the NK-model. So, our simulator can be used for experiments on the NK-model.

As a concrete example, suppose that we have $K=2$, a locus $x_1$ and another locus $x_2$ with which it interacts, and the fitness contribution

\[f_1(0,0) = 3\]
\[f_1(0,1) = 5\]
\[f_1(1,0) = 7\]
\[f_1(1,1) = 9\]

Then, this can be represented in the simulator as

%%TC:ignore
\begin{minted}{python}
def fitFunc(x,y):
    if x == 0 and y == 0:
        return 3
    if x == 0 and y == 1:
        return 5
    if x == 1 and y == 0:
        return 7
    if x == 1 and y == 1:
        return 9
ConstraintVCSP([x_1,x_2], (fitFunc, 9))
\end{minted}
%%TC:endignore

\subsection{Colouring the map of Romania}

In this subsection, we present the results of running experiments on our simulator using the problem of colouring the map of Romania, introduced in Section \ref{vcspBackground}.

For encoding the problem we use the Binary Model constraints. With this formulation, we define the $2$ - colouring problem for the map of Romania, where we prefer $C$ to be $true$ (red) and $D$ to be $false$ (blue) in the following way (with unit weight for different values for the neighbours):

%%TC:ignore
\begin{minted}{python}
initial = [0] * 8
# C - 0, MA - 1, MO - 2, DO - 3, MU - 4, O - 5, B - 6, T - 7
probability = 0.01 * 8
rounds = 50
orgNum = 100
cons1 = ConstraintBinaryModelUnary(0, 1)
cons2 = ConstraintBinaryModelUnary(3, -1)
cons3 = ConstraintBinaryModelBinaryDifferent([0,7],[1,1]) #(C != T)
cons4 = ConstraintBinaryModelBinaryDifferent([0,1],[1,1]) #(C != MA)
cons5 = ConstraintBinaryModelBinaryDifferent([1,7],[1,1]) #(MA != T)
cons6 = ConstraintBinaryModelBinaryDifferent([1,2],[1,1]) #(MA != MO)
cons7 = ConstraintBinaryModelBinaryDifferent([7,2],[1,1]) #(T != MO)
cons8 = ConstraintBinaryModelBinaryDifferent([2,3],[1,1]) #(MO != D)
cons9 = ConstraintBinaryModelBinaryDifferent([4,3],[1,1]) #(MU != D)
cons10 = ConstraintBinaryModelBinaryDifferent([4,7],[1,1]) #(MU != T)
cons11 = ConstraintBinaryModelBinaryDifferent([4,5],[1,1]) #(MU != O)
cons12 = ConstraintBinaryModelBinaryDifferent([7,5],[1,1]) #(T != O)
cons13 = ConstraintBinaryModelBinaryDifferent([6,5],[1,1]) #(B != O)
cons14 = ConstraintBinaryModelBinaryDifferent([6,7],[1,1]) #(B != T)
cons15 = ConstraintBinaryModelBinaryDifferent([6,0],[1,1]) #(B != C)
constraints = [cons1, cons2,cons3,cons4,cons5,
    cons6,cons7,cons8,cons9,cons10,cons11,cons12,cons13,cons14,cons15]

mySim = Simulator(2, initial, probability, rounds,
    orgNum, constraints, None, 1, 1)
#the first value represents the type 2 - binary constraints
#no domain is required as it is boolean by default
mySim.run()
mySim.printStatistics()
mySim.printLocalStatistics()
\end{minted}
%%TC:endignore

%%TC:ignore
\begin{figure}
\begin{minipage}{.5\textwidth}
  \includegraphics[scale = 0.68]{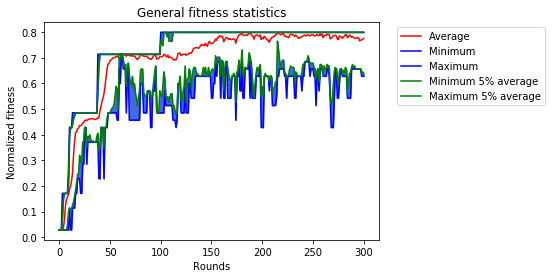}
\end{minipage}
\caption{Plot of the fitness statistics for the Romania map colouring}
\label{fig:fig17}
\end{figure}
%%TC:endignore

%%TC:ignore
\begin{figure}
\centering
\begin{minipage}{.5\textwidth}
  \centering
  \includegraphics[scale = 0.53]{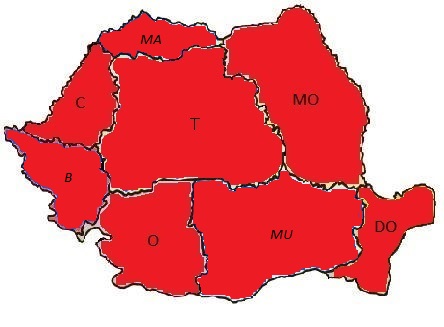}
\end{minipage}%
\begin{minipage}{.5\textwidth}
  \centering
  \includegraphics[scale = 0.53]{Romania4.jpg}
\end{minipage}
\caption{The initial colouring of the map of Romania in our simulator (left) and the final one (right).}
\label{fig:fig21}
\end{figure}
%%TC:endignore

The results are shown in Figure \ref{fig:fig17}. For this setting, we cannot satisfy all the constraints, so we peak at about $0.8$. Figure \ref{fig:fig21} shows the initial colouring and the final one.

\newpage

\section{Theoretical results} \label{model}

In this section, we propose a mathematical model that encapsulates the smooth landscape in Section \ref{smoothLandscape} and compare the results with the ones from the simulator, showing that they arrive at the same conclusions. We use some ideas from the discrete Markov chains theory (Appendix C). A similar approach, but for a different simulation of evolution and a different result is used by \cite{nichol2013markov}.

We model the number of organisms at each fitness level after a number of rounds $k$. We have  $n + 1$ fitness levels, where $n$ is the size of the genome, corresponding with a fitness of $0, 1, .. n$. Thus, we can denote the proportion by a (row) vector of size $n + 1$, $w_k$. The initial configuration will correspond to all organisms having a fitness of $0$, where $m$ is the total number of organisms:

\[w_0 = [m, 0, 0, ..., 0]\]

%%TC:ignore
\begin{figure}
\begin{minipage}{.5\textwidth}
  \includegraphics[scale = 0.6]{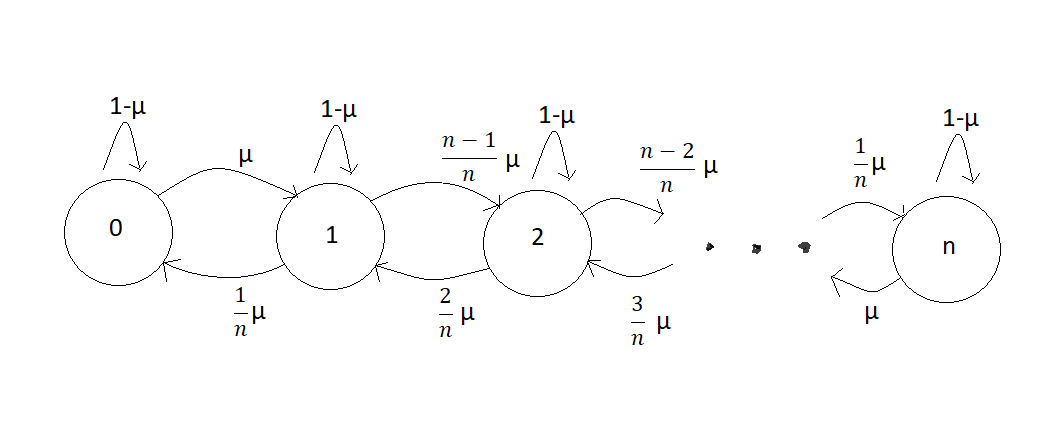}
\end{minipage}
\caption{A representation of the smooth Markov chain model}
\label{fig:fig18}
\end{figure}
%%TC:endignore

If we denote the mutation probability by $\mu$, then, for an organism with a fitness of $i$, there is a probability of $1 - \mu$ of not suffering any mutation. In addition, there is a probability of $\mu\frac{n - i}{n}$ to get to a fitness level of $i + 1$ (as we already have $i$ bits set to $1$, there is a chance of $\frac{n - i}{n}$ to mutate one of the $0\text{s}$), which means that there is a probability of $\mu\frac{i}{n}$ to mutate to a fitness level of $i - 1$ (Figure \ref{fig:fig18}). Thus, we obtain a probability matrix $P$, of size $(n + 1) \times (n + 1)$:

\[P_{i,i+1} = \mu\frac{n - i}{n}, \text{ for } i = 0, 1, ... , n - 1\]

\[P_{i,i-1} = \mu\frac{i}{n}, \text{ for } i = 1, 2, ... , n\]

\[P_{i,i} = 1 - \mu, \text{ for } i = 0, 1, ... , n\]

Then, $w_kP$ will be the (expected) population distribution after mutating. However, this does not take into account the reproduction of the organisms based on the fitness levels. So, we need to supply the expected number of children at each fitness level. This is equal to the $fitness + 1$, as it is is defined as a sampling from a Poisson distribution with this parameter. We represent this as a $(n + 1) \times (n + 1)$ diagonal matrix $F$:

\[F_{i,i} = i + 1, \text{ for } i = 0, 1, ... , n\]
\[F_{i,j} = 0, \text{ for } 0 \leq i \neq j \leq n\]

Then, the population at step $k + 1$ will be

\[w_{k + 1} = w_kFP\]

This respects the order of operations in our simulator, where we first sample the number of offspring (matrix $F$), then we apply the mutations (matrix $P$).

If we set $A = FP$, the equation becomes:

\[w_{k + 1} = w_kA\]

By deriving the recursion, we obtain:

\[w_k = w_0A^k \]

This gives us the population after $k$ steps. However, in this vector, the total population is greater than $m$, as we don't take into account the hypergeometric distribution sampling that happens in the simulation. We can do this by applying a regularisation coefficient, such that the final result $w_k^F$ equals the expected value of the hypergeometric distribution. Thus, if we define $M$ to be the number of organisms in $w_k$, we get

\[w_k^F = \frac{m}{M}w_k = \frac{m}{M}w_0A^k\]

If we define the vector \textbf{1} as the row vector of size $n + 1$ made of $1\text{s}$, we can compute $M$ as follows:

\[M = \sum_{j=0}^{n} w_{k,j} = \textbf{1} \cdot w_k \]

Thus,

\[w_k^F = \frac{m}{\textbf{1} \cdot w_k}w_k = \frac{m}{\textbf{1} \cdot w_0A^k}w_0A^k = \frac{m}{w_0A^k\textbf{1}^T}w_0A^k\]

Instead of computing the distribution of the number of organisms, we can calculate the distribution of the proportion of the population in each fitness level (as a vector $v_k$). This is done in the same way as above, but we consider $m = 1$. Thus,

\[v_0 = [1, 0, 0, ..., 0]\]

\[v_{k + 1} = v_kA\]

\[v_k = v_0A^k \text{ (1)}\]

\[v_k^F  = \frac{1}{v_0A^k\textbf{1}^T}v_0A^k \text{ (2)}\]

The equation in $(1)$ can be solved (for $k \rightarrow \infty$), and then plugged into $(2)$. The same can be done if we find a stationary distribution, for some $k$ (Appendix C).

Thus, we obtained a closed form of the equation, obtaining a formula for the proportion of the population after $k$ rounds, given the initial distribution $v_0$, the probability matrix $P$ and the matrix of children per fitness level $F$.

The above equation is closely related to the discrete replicator-mutator equation, which appears in the literature (\cite{page2002unifying}, \cite{harper2012stability}).

The average fitness at step $k$, $\overline{f_k}$, can be computed by the formula:

\[\overline{f_k} = [1, 2, .., n + 1] \cdot v_k = v_kF\textbf{1}^T\]

The average fitness of the fitter mutants at step $k$ is computed as follows:

\[\overline{f_k^{N+}} = [2, 3, .., n + 1, n + 1] \cdot v_k\]

Observe that for each level we consider the mutants of the next level, while for the last level we keep the same mutants.

Thus, the selection coefficient at step $k$ is:

\[s_k = \frac{\overline{f_k^{N+}} - \overline{f_k}}{\overline{f_k}}\]
\[s_k = \frac{v_k \cdot [2, 3, .., n + 1, n + 1] - v_k \cdot [1, 2, .., n, n + 1]}{v_k \cdot [1, 2, .., n + 1]}\]
\[s_k = \frac{v_k \cdot [1, 1, .., 1, 0]}{v_k \cdot [1, 2, .., n + 1]}\]

If we write $v_k$ in terms of $v_{k-1}$, we obtain:

\[s_k = \frac{\frac{1}{v_{k-1}F\textbf{1}^T} * (v_{k-1}FP)[1, 1, .., 1, 0]^T}{\frac{1}{v_{k-1}F\textbf{1}^T} * (v_{k-1}FP)[1, 2, .., n + 1]^T}\]
\[s_k = \frac{(v_{k-1}FP)[1, 1, .., 1, 0]^T}{(v_{k-1}FP)[1, 2, .., n + 1]^T}\]

We can break the matrix $P$ in $3$ parts. A diagonal part, which is the identity matrix $I$, an upper diagonal matrix that defines the probabilities to get to a higher fitness, $P^+$, and a lower diagonal matrix that defines the probabilities to get to a lower fitness, $P^-$.

\[P^+_{i,i+1} = \mu\frac{n - i}{n}, \text{ for } i = 0, 1, ... , n - 1\]
\[P^+_{i,j} = 0, \text{ otherwise }\]
\[P^-_{i,i-1} = \mu\frac{i}{n}, \text{ for } i = 1, 2, ... , n\]
\[P^-_{i,j} = 0, \text{ otherwise }\]

Then, $P$ can be written as follows:

\[P = (1 - \mu) I + \mu P^+ + \mu P^-\]

If we assume a very small mutation probability $\mu \rightarrow 0$, then $P \rightarrow I$. So, the selection coefficient becomes:

\[s_k = \frac{(v_{k-1}F)[1, 1, .., 1, 0]^T}{(v_{k-1}F)[1, 2, .., n + 1]^T}\]

We can generalise the above formulas for multiple fitness levels by defining the matrix $F$ as:

\[F_{i,i} = g(i), \text{ for } i = 0, 1, ... , n\]
\[F_{i,j} = 0, \text{ for } 0 \leq i \neq j \leq n\]

For some function $g$ that gives us the fitness levels.

%%TC:ignore
\begin{figure}
\centering
\begin{minipage}{.5\textwidth}
  \centering
  \includegraphics[scale = 0.44]{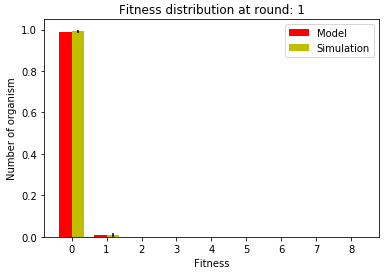}
\end{minipage}%
\begin{minipage}{.5\textwidth}
  \centering
  \includegraphics[scale = 0.44]{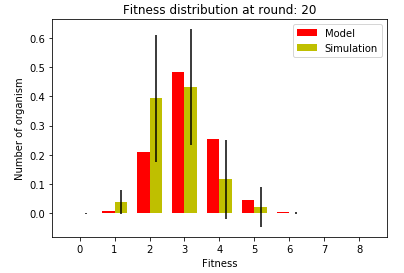}
\end{minipage}
\begin{minipage}{.5\textwidth}
  \centering
  \includegraphics[scale = 0.44]{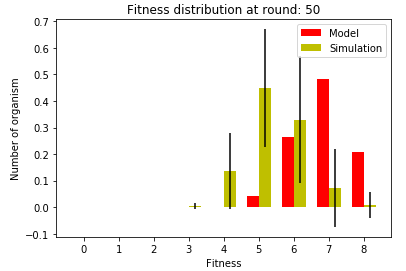}
\end{minipage}%
\begin{minipage}{.5\textwidth}
  \centering
  \includegraphics[scale = 0.44]{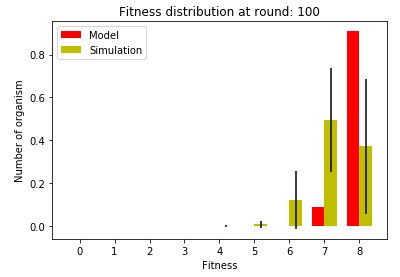}
\end{minipage}
\begin{minipage}{.5\textwidth}
  \centering
  \includegraphics[scale = 0.44]{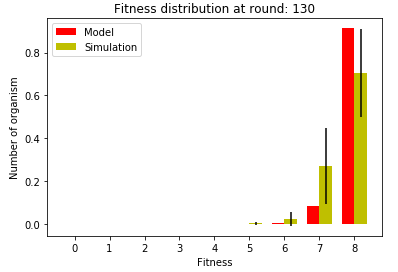}
\end{minipage}%
\begin{minipage}{.5\textwidth}
  \centering
  \includegraphics[scale = 0.44]{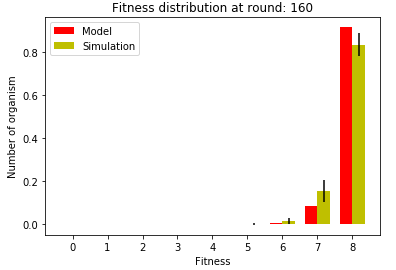}
\end{minipage}
\begin{minipage}{.5\textwidth}
  \centering
  \includegraphics[scale = 0.44]{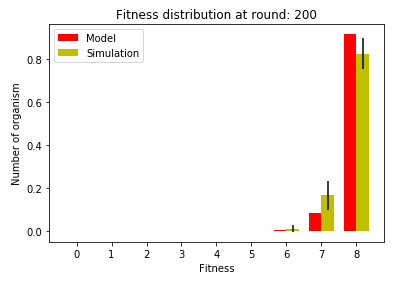}
\end{minipage}%
\begin{minipage}{.5\textwidth}
  \centering
  \includegraphics[scale = 0.44]{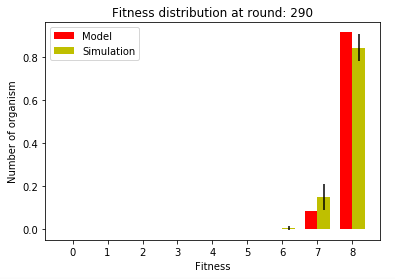}
\end{minipage}
\caption{Comparisons of the distribution of organisms by fitness in the model and the simulator for a selection of rounds }
\label{fig:fig8}
\end{figure}
%%TC:endignore

We can compare the above model with the simulator. We compute the distributions at each round (for our smooth landscape problem in Section $4.1$) using both the model and $50$ repetitions of the simulator, for which we show the average distribution and the standard deviation. The results (Figure \ref{fig:fig8}) show that in general there is a close match between the results of the model and the simulator. However, before the convergence happens, the results tend to differ more, while at the same time observing that the data for the simulation has a very high variance, which comes from the great number of possible mutations that can happen before the population stabilises. 

%%TC:ignore
\begin{figure}
\begin{minipage}{.5\textwidth}
  \includegraphics[scale = 0.35]{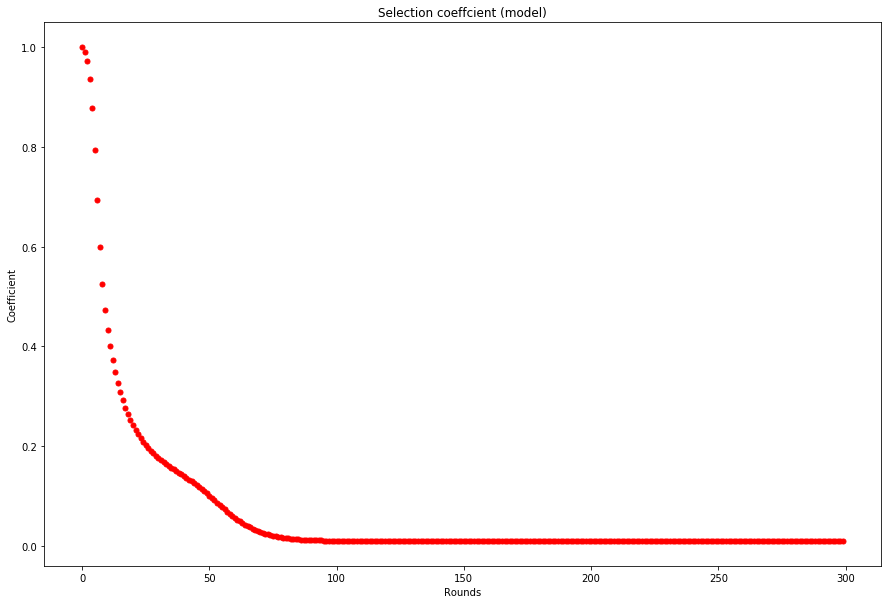}
\end{minipage}
\caption{The plot of the selection coefficient computed from the model for each round.}
\label{fig:fig3}
\end{figure}
%%TC:endignore

Using the above formulas, we can also compute the selection coefficient for the model. As expected, we observe an exponential decay (Figure \ref{fig:fig3}), similar with what is obtained in the simulations (Figure \ref{fig:fig6}).

%%TC:ignore
\begin{figure}
\begin{minipage}{.5\textwidth}
  \includegraphics[scale = 0.75]{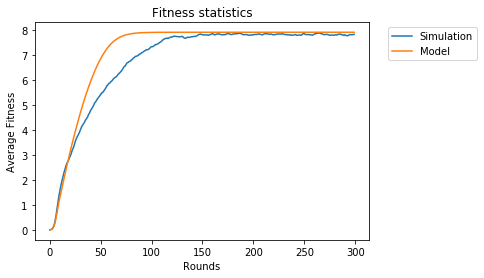}
\end{minipage}
\caption{Comparison of the average fitness for the simulation and the model}
\label{fig:fig9}
\end{figure}
%%TC:endignore

From a further statistic, which shows the average fitness in both the model and the simulation (Figure \ref{fig:fig9}), we observe that the two values converge to the same result when the population stabilises, so we can conclude that our model is a good one for predicting the outcome of evolution on smooth landscapes.

Finally, we consider what happens when we try to generalise our model for other landscapes. If we have a genome of length $n$, we can have $2^n$ gene configurations, so there is an exponential number of fitness levels, which gives an exponential number of nodes in our chain. This means that we are no longer guaranteed to reach the equilibrium after a reasonable (polynomial) time. While our landscape can still be easy to navigate, depending on the edges along the fitness graph, it can also have a more intricate geometry, which will imply going for a very long path, as we have observed on our hard semismooth landscape (Section \ref{semiSmooth}.

\newpage

\section{Conclusions} \label{conclsuions}

\subsection{Summary of the work}

As planned from the beginning in the project description, we managed to address three main aspects: a programming aspect, an experimental aspect and a theoretical aspect.

From the programming perspective, we built a new evolution simulator that is able to run experiments using various types of VCSPs, which is something new for the field. We used a simple simulation method, by keeping a (fixed-size) population that reproduces in each round based on the fitness of the individuals. We used a series of probability distributions from which we sample the next generation.

Along with the main simulation, our program computes a series of global and local statistics that are used to interpret the evolution of the population (the most important being the fitness, selection coefficient and distance from the peak).

From the experimental perspective, we ran a series of tests on the three types of fitness landscapes identified in \cite{artem2}. Those tests demonstrate that the simulation proceeds as expected from the previous studies. However, we also obtained more accurate information about the expected convergence times than was previously known. Furthermore, the experiments for a candidate hard semismooth landscape were particularly surprising, because they showed that, although the landscape is only proved to be hard on strict fittest-mutant SSWM, it remains hard on our simulator (implementing fitter-mutant SSWM), which is believed to be closer to a naturally occurring dynamics. Finally, we showed that the NK-model can be encoded in our simulator.

From the theoretical perspective, we presented a (deterministic) mathematical model for the smooth landscapes and compared the results with the (stochastic) simulator.

\subsection{Further work}

As this is the first biological evolution simulator for VCSP problems, it opens a whole set of problems to be answered, and other extensions can be made.

Firstly, we can run experiments on hard rugged landscapes in a similar way as our example for the semismooth landscape. This would imply finding a hard fitness landscape and encoding it as VCSP.

Secondly, we can generalise the recursive fitness function for the hard semismooth landscape (from \cite{artem2}) and consider more padding. Another approach would be to encode the fitness landscape that is hard for random fitter mutant SSWM, which can be found in \cite{matouvsek2006random}.

Finally, work can be done on the simulator by looking for more optimisation or using parallelism, as suggested in Section \ref{challenge}.

\subsection{Critical review}

Overall, the project reached its aims of producing an evolution simulator, running simulations and getting theoretical conclusions from them. The most important aspects learned from this project include reading academic background about a current need (in our case a way to experimentally confirm some theoretical work) and finding an appropriate solution (the simulator). This implied developing a design that will offer a balance between efficiency and usability and writing code in a new computer language using new frameworks to which I had little previous exposure. Using this tool I learned about running simulations and computing relevant statistics from the data, as well as explaining the results by developing a sound theory. Finally, everything had to be put together into an appropriate report.

An important step for those kind of programming projects is to identify early as many needed components as possible, and to follow a modular approach, as it can be hard to later add new, critical features. I feel that thinking more thoroughly about these aspects in the early stages could have saved some development time that was later needed to make the design more flexible. 

\newpage

\section{Appendices}

\subsection{Appendix A}

This subsection presents some useful concepts and results from Statistics. The first part is based on \cite{statisticsOx}.

\begin{defn} 
A \textbf{random sample} of size $n$ is a collection of items (represented as random variables, e.g. $X_1, X_2, ..., X_n$) drawn independently from some set of items (population) with some probability distribution. 
\end{defn}

\begin{defn}
For a random variable $X$, we denote the \textbf{expectation} of $X$, or its mean, as $\mathbf{E}[X]$. If $X$ is known from the context, this can also be denoted as $\mu$. 
\end{defn}

\begin{defn}
For a random variable $X$, we define the \textbf{variance} of $X$ as $var(X) = \mathbf{E}[(X - \mu)^2]$. This is also denoted as $\sigma^2$, while $\sigma$ is known as the \textbf{standard deviation}.
\end{defn}

\begin{defn}
For two random variables $X$, $Y$, we define the \textbf{covariance} of $X$ and $Y$ as $cov(X, Y) = \mathbf{E}[(X - \mu)(Y - \mu)]$, also denoted as $\sigma_{X,Y}$.
\end{defn}

\begin{defn} 
For a given random sample $X_1, X_2, ..., X_n$, we define the \textbf{sample mean} as
\[\overline{X} = \frac{1}{n}\sum_{i=1}^nX_i\]
\end{defn}

We use the sample mean extensively in Section \ref{results} to compute the average values for different measure of our sampled population. From a statistical point of view, our populations of organisms are considered to be a random sample from the true distribution of organisms.

\begin{defn} 
For some given random sample $X_1, X_2, ..., X_n$, we define the \textbf{sample variance} as:
\[S^2 = \frac{1}{n-1}\sum_{i=1}^n(X_i-\overline{X})^2 \]
The \textbf{sample standard deviation} is $S$.
\end{defn}

The two statistics above are examples of estimators (for $\mu$ and $\sigma$ respectively). We say that an estimator is biased if the difference between its expected value and true value is non zero.  Note that in the above definition, $n-1$ is used instead of $n$ in order to have an unbiased estimator.

In Section \ref{localStatsClass}, we use \textbf{propagation of uncertainty} to compute the standard error for the selection coefficicent. We describe this technique based on \cite{ku1966notes}.

The formula for the selection coefficient (as in Section \ref{bioBack}) is 

\[s_k = \frac{\overline{f^{N+}_k} - \overline{f_k}}{\overline{f_k}}\]

We denote

\[a = \overline{f^{N+}_k}, b = \overline{f_k}\]

So, we have,

\[\sigma_a = \frac{\sigma_{f_k^{N+}}}{\sqrt{n}}\]
\[\sigma_b = \frac{\sigma_{f_k}}{\sqrt{n}}\]
\[\sigma_{a,b} = cov(\overline{f^{N+}_k},\overline{f_k}) = \frac{1}{n}cov(f^{N+}_k,f_k)\]

Thus,

\[s(a,b) = \frac{a - b}{b} = \frac{a}{b} - 1\]

By the first order Taylor expansion, we obtain

\[s \approx s^0 + \frac{\partial s}{\partial a}a + \frac{\partial s}{\partial b}b\]

From this, we derive the variance of $s$:

\[\sigma_s^2 \approx \Big|\frac{\partial s}{\partial a}\Big|^2\sigma_a^2 + \Big|\frac{\partial s}{\partial b}\Big|^2\sigma_b^2 + 2\frac{\partial s}{\partial a} \frac{\partial s}{\partial b} \sigma_{a,b}\]

Then, we compute the partial derivatives:

\[\frac{\partial s}{\partial a} = \frac{1}{b}\]
\[\frac{\partial s}{\partial b} = -\frac{a}{b^2}\]

Finally, the variance of $s$ is:

\[\sigma_s^2 \approx \frac{a^2}{b^2} \bigg( \Big( \frac{\sigma_a}{a} \Big)^2 + \Big( \frac{\sigma_b}{b} \Big)^2 - 2 \frac{\sigma_{a,b}}{ab} \bigg)\]

So the standard deviation is obtained by taking the square root from the above formula.

\subsection{Appendix B}

This section introduces some probability distributions that are used in our simulators. The presentation is based on \cite{probOx} and \cite{shanmugam2015statistics}.

\subsubsection{The Bernoulli Distribution}

\begin{defn} 
A random variable $X$ has a Bernoulli distribution with parameter $ 0 \leq p \leq 1 $ if,
\[\mathbb{P}(X=0) = 1 - p\]
\[\mathbb{P}(X=1) = p\]
\end{defn}

The expectation (mean) of $X$ is $p$, while the variance is $p(p-1)$.

\subsubsection{The Poisson Distribution}

\begin{defn} 
We say that a random variable $X$ has a Poisson distribution with parameter $\lambda \geq 0$ if,
\[\mathbb{P}(X=k) = \frac{\lambda^ke^{-\lambda}}{k!}, k \geq 0\]
\end{defn}

Both the expectation and the variance of a random variable having a Poisson distribution with parameter $\lambda$ is $\lambda$. This is extremely useful, as we use the parameter of the Poisson distribution to be the fitness of the organism when sampling its offspring. This results in having the expected number of children equal to the fitness.

\subsubsection{The Hypergeometric Distribution}

\begin{defn} 
We say that a random variable $X$ has a (bivariate) hypergeometric distribution with parameters $N \geq 0$, $0 \leq k \leq N$ and $0 \leq n \leq N$ if,
\[\mathbb{P}(X=x) =\frac{{k\choose x}{N - k\choose n - x}}{{N \choose n}}, 0 \leq x \leq \min{(n,k)}\]
\end{defn}

This corresponds with having $N$ objects of two indistinguishable types ($k$ of the first type and $N-k$ of the second). If we sample $n$ times without replacement from the collection of objects, the above equation gives the number $x$ of the first type obtained.

The mean is equal to $n\frac{k}{N}$, while the variance is $n\frac{k}{N}\frac{N-k}{N}\frac{N-n}{N-1}$.

\begin{defn} 
For some $c \geq 0$, we say that a random variable $X_i$, with $0 \leq i \leq c$ has a multivariate hypergeometric distribution with parameters $k_j \geq 0$ ($j \leq i \leq c$), $N = \sum_{j=1}^c k_i$,  and $0 \leq n \leq N$ if,
\[\mathbb{P}(X_i=x_i) =\frac{\prod_{i=1}^c{k_i \choose x_i}}{{N \choose n}}, 0 \leq x_i \leq \min{(n,k_i)}\]
\end{defn}

This corresponds with having $c$ different types of indistinguishable objects (rather than just $2$), each one having $k_i$ members. From this, we choose $n$ objects without replacement. Then, the above distribution gives the probability of having $x_i$ objects of type $i$. This is exactly like our case from the simulator.

For $X_i$, the mean is equal to $n\frac{k_i}{N}$, while the variance is $n\frac{k_i}{N}\frac{N-k_i}{N}\frac{N-n}{N-1}$. The covariance between $X_i$ and $X_j$ is $-\frac{nk_ik_j}{N^2}\frac{N-n}{N-1}$.

\subsection{Appendix C}

We first present some ideas about the Markov chains. Next, we show techniques for finding the limit in Markov chains and other sequences based on matrix power, using the eigenvalues and eigenvectors. These are used in our model from Section \ref{model}. This presentation is based on \cite{probOx2} and \cite{linearOx}.

\subsubsection{Markov chains}

\begin{defn} 
Let $I$ be the state space and let $X = (X_1, X_2, ...,)$ be a (infinite) sequence of random variables with values from $I$. We say that the above process is a (discrete-time) \textbf{Markov Chain} if for all $n \geq 0$ and $i_0, i_1, ..., i_{n+1} \in I$,
\[\mathbb{P}(X_{n+1}=i_{n+1}|X_n=i_n,...,X_0=i_0)=\mathbb{P}(X_{n+1}=i_{n+1}|X_n=i_n)\]
\end{defn}

The above relation means that the probability distribution depends only on the current state.

If $\mathbb{P}(X_{n+1}=j|X_n=i)$ depends only on $i$ and $j$, then the chain is homogeneous, so we write $p_{i,j} = \mathbb{P}(X_{n+1}=j|X_n=i)$. The quantities $p_{i,j}$ can be represented as a matrix, known as the transition matrix $P = (p_{i,j})$, in which the values are non-negative, and the sum along the rows is $1$. This can also be represented through a diagram. The distribution at each step $n$ is denoted by a vector of probabilities $x_n$, which sums to $1$. Thus, for defining a Markov chain, we have to give the initial distribution $x_0$ and the transition matrix $P$. Then, the distribution is defined recursively as $x_{n+1} = x_nA$. So, at step $n$, we have $x_n = x_0A^n$. We can consider what happens when $n \rightarrow \infty$ by defining $x_{\infty} = lim_{n\to\infty}(x_0A^n)$. Finally, we say that a distribution vector $\pi$ is stationary if $\pi = \pi A$. Those two cases are analysed in the next subsection, using eigenvalues and eigenvectors.

\subsubsection{Some concepts of linear algebra}

\begin{defn}
For some $N \times N$ matrix $A$, a non-zero vector $u$ and a scalar $\lambda$, if $Au = \lambda u$, we say that $\lambda$ is an \textbf{eigenvalue} and $u$ the corresponding \textbf{eigenvector} of $A$.
\end{defn}

We can compute the eigenvalues (if they exist), by solving the equation $\det(A-\lambda I) = 0$. From this, we find $u$ by finding a non-zero vector that solves $(A-\lambda I)u = 0$.

If $A$ has $N$ distinct eigenvalues, then there exist a diagonal matrix (of eigenvalues) $D$ and an invertible matrix $S$ such that $D = S^{-1} A S$. We say that $A$ is diagonalised. Then, as $A = SDS^{-1}$, we have (by induction on $n$) that $A^n = SD^nS^{-1}$. If the eigenvalues are between $0$ and $1$, then for $n \rightarrow \infty$, $D^n$ converges to a finite valued matrix, so $A^n$ is finite. Finally, note that the stationary distribution for a Markov chain, from the previous subsection, is actually an eigenvalue of the transition matrix.

We can use those techniques to solve the equations for our model from Section \ref{model}.

\subsection{Appendix D} \label{appD}

Here, we give a detailed presentation of the components of the simulator together with some relevant parts of the code.

\subsubsection{Importing the module and adding input}

The module can be imported using

%%TC:ignore
\begin{minted}{python}
from Simulator import *
\end{minted}
%%TC:endignore

The input is represented by the number of organisms (\textit{orgNum}), the number of rounds for which the simulator will run (\textit{rounds}), the probability of a gene mutation (\textit{probability}), the initial genome (\textit{initial}), the array of domains of each variable (\textit{domains}) and the constraints, represented as an array of constraint objects(\textit{constraints}). The problem type is also passed to the simulator, corresponding to the type of the constrains used (\textit{probType}).

To run the simulation, create an instance of it by passing the input and calling the \textit{run} method. The statistics are showed using the methods \textit{Statistics} and \textit{LocalStatistics}.

\subsubsection{The \textit{Simulator} class} \label{simulator}

The \textit{Simulator} class will initialise the variables with the input given and will create the \textit{Population} class. A call to \textit{run} will run the simulator for the required number of rounds, computing the next generation after each run. The methods \textit{printLocalStatistics} and \textit{printStatistics} will call the respective functionalities and print the graphs. The \textit{probType} variable will give us the type of the problem solved. For SAT and binary constraint  (types $1$ and $2$), it will also define the only domain possible. The \textit{fitOffset} variable defines a fitness offset that will be added when computing the fitness (usually $1$, so that $fitness \geq 1$).

The \textit{mutType} variable encodes the way we do mutations. The \textit{maxGenome} is used in statistics to compute the distance to the maximum possible genome during the simulation.

%%TC:ignore
\begin{minted}{python}
class Simulator:  
    def __init__(self,
    	     probType,		 
                 initial,         
                 probability,     
                 rounds,          
                 orgNum,          
                 constraints,     
                 domains,         
                 fitOffset,
                 maxGenome = None):		 
        self._rounds = rounds
        self._constraints = constraints
        self._genLength = len(initial)
        self._probType = probType
        self._distrib = []
        if(probType == 1 or probType == 2):
        	domains = []
        	for i in range (0, len(initial)):
        		domains.append([0,1])     
        self._population = Population(
            orgNum, constraints, probability, initial, domains,
                fitOffset)
    
    def run(self):
        for i in range (0, self._rounds):
            self._population.nextGeneration()    
    def printLocalStatistics(self):
    	self._population.getLocalStats().plotStats()
    	
    def printStatistics(self):
    	self._population.getStats().plotStats()
\end{minted}
%%TC:endignore

\subsubsection{The \textit{Population} class} \label{population}

The \textit{Population} class represents the total population of organisms, which is described as an array with members from the \textit{Organism} class. It also constructs and initialises the objects representing the statistics.

The \textit{nextGeneration} method computes the next population after the current round. It generates the pool of offspring, from which only \textit{orgNum} will remain in the next round. This is done by sampling from the multivariate hypergeometric distribution. A dictionary of labels is used to represent each new genome that is generated from mutations. After the sampling is finished, the labels are used to construct the new genome. Then, the mutations for each organism are triggered by calling the \textit{mutate} function.

%%TC:ignore
\begin{minted}{python}
def nextGeneration(self):
    nextGenPool = []
    self._round += 1
    dct = {}                                                  
    for i in range(0, self._orgNum):
        nextGenPool.append(self. _organisms[i].offspring(
            totalFitness, self._orgNum, i, dct))       
    
    sumNextPool = 0
    
    for (num, label) in nextGenPool:                      
    	sumNextPool += num
    totalRemaining = self._orgNum                                
    self._organisms = []
    
    for (num, label) in nextGenPool[:-1]:
        if totalRemaining <= 0:
            currNum = 0
        else:
            currNum = np.random.hypergeometric(
                num, sumNextPool - num, totalRemaining)  
        sumNextPool -= num
        totalRemaining -= currNum
        for i in range (0, currNum):
            org = dct[label]
            org.mutate() 
            self._organisms.append(org)
    
    (num, label) = nextGenPool[-1]
    
    for i in range (0, totalRemaining):   
        org = dct[label]
        org.mutate() 
    	self._organisms.append(org)

    self._stats.updateStats()                                
    self._localStats.updateStats()
\end{minted}
%%TC:endignore

\subsubsection{The \textit{Organism} class} \label{organism}

The \textit{Organism} class represents an organism from the population.

The method \textit{computeFitness} computes the current value of the fitness function by adding the reward from each clause, as returned by the \textit{evaluate} function, together with the offset.

%%TC:ignore
\begin{minted}{python}
def _computeFitness(self):
    solvedConstraints = 0
    constraintNum = len(self._constraints)
    for clause in self._constraints :
        solvedConstraints += clause.evaluate(self._genome,
            self._domains)
    return solvedConstraints + self._fitOffset 
\end{minted}
%%TC:endignore

The method \textit{mutate} computes a mutation of the current genome. Based on the given argument \textit{mutType}, we can have two types of mutations. The first type involves iterating through the genes / variables and sampling from a Bernoulli distribution (binomial with \textit{n} of value 1) with the initially given probability of mutation. If the value is $1$, the gene value is changed. Otherwise, it remains the same. If the value is changed, a new sampling is done to choose which value in the domain will be used. The offset is added, guaranteeing that a new value is chosen. The genome and the fitness variables are then updated accordingly. Thus, we can have more than a single gene to suffer a mutation during a \textit{mutate} sequence.

For the second type, we will have at most a single mutating gene. In this way, we first decide if we mutate or not using the Bernoulli distribution. If we decide to mutate, we randomly choose a gene for which we apply the same mutation procedure as above.

%%TC:ignore
\begin{minted}{python}
def mutate(self):     
    if self._mutType == 0:
        varNum = len(self._genome)
        for i in range(0, varNum) :
            doMutation = np.random.binomial(1, self._prob, 1)
            if doMutation == 1:
                domain = self._domains[i]
                x = 1
                domLen = len(domain)
                if domLen > 2:                                      
                    x = np.random.randint(1, domLen)                
                self._genome[i] = (x + self._genome[i]) % domLen
    else:
        doMutation = np.random.binomial(1, self._prob, 1) 
        varNum = len(self._genome)
        if doMutation == 1: 
            i = np.random.randint(0, varNum)
            domain = self._domains[i]
            x = 1
            domLen = len(domain)
            if domLen > 2: 
                x = np.random.randint(1, domLen) 
            self._genome[i] = (x + self._genome[i]) % domLen
    self._fitness = self._computeFitness()
\end{minted}
%%TC:endignore

The method \textit{offspring} returns a list of children based on the current fitness value. This is done by sampling from the Poisson distribution, with a mean equal to the $fitness - 1$ (we remove the artificially added $1$). This value is again incremented, so that each organism will have at least one child. We keep a dictionary from a label to the genome. This is used by the \textit{Population} class to construct the next generation.

%%TC:ignore
\begin{minted}{python}
def offspring(self, totalFitness, orgNum, label, dct):                    
    myFitness = self._fitness               
    s = int(np.random.poisson(myFitness - 1, 1) + 1)    
    dct[label] = Organism(self._genome.copy(), self._env, self._prob);
    return (s, label)  
\end{minted}
%%TC:endignore

The method \textit{computeMutants} returns all the possible mutants of an organism at Hamming distance $1$.

%%TC:ignore
\begin{minted}{python}
def computeMutants(self) : 
    res = []
    l = len(self._genome)
    for i in range (0, l) :
        domain = self._domains[i]
        domLen = len(domain)
        for off in range (1, domLen):
            newGenome = self._genome.copy()
            newGenome[i] = (newGenome[i] + off) % domLen
            res.append(Organism(newGenome, self._constraints,
                self._prob, self._domains, self._fitOffset))
    return res
\end{minted}
%%TC:endignore

\subsubsection{The \textit{Constraint} class and its subclasses} \label{constraints}

The \textit{Constraint} class is an abstract class that represents the structure of a constraint. It is subclassed by other classes for making concrete definitions of constraints.

The \textit{evaluate} method returns the reward obtained from a clause given the current variables and the domain, while the \textit{getWeight} method returns the weight of the constraint.

%%TC:ignore
\begin{minted}{python}
class Constraint:
	_metaclass_ = ABCMeta
	
	@abstractmethod
	def evaluate (self,variables): pass
	
	@abstractmethod
	def getWeight(self): pass
\end{minted}
%%TC:endignore

The most general \textbf{VCSP model} is implemented by the class \textit{ConstraintVCSP}. Each constraint is given by the constrained variables and a function that will return the reward based on the variable values. The user supplies a maximum reward value that is used for computing the relative fitness.

%%TC:ignore
\begin{minted}{python}
class ConstraintVCSP(Constraint):					 
	def __init__(self, elems, weightInfo):   	 
		self._elems = elems 					 
		self._weightFunction = weightInfo[0]      
		self._maxWeight = weightInfo[1]         
	def evaluate(self, variables, domains):
		args = []
		for elem in self._elems:
			args.append(domains[elem][variables[elem]])
		#the values of the variables are computed and 
		#passed to the weight function
		return self._weightFunction(args)
	def getWeight(self):
		return self._maxWeight
\end{minted}
%%TC:endignore

The \textbf{weighted SAT model} is implemented by the \textit{ConstraintSat} class. Each constraint represents an $OR$ clause of literals. The constraint is satisfied if at least a single literal is $true$. A constraint is defined by a list of elements, each representing a variable in the genome and a value, representing the weight for satisfying the clause. A positive value means that the literal is positive, while a negative value defines a negated variable. The number in absolute value represents the position in the genome (starting from $1$). The evaluate function computes the value of the clause, given the values for the variables. The domain is always the boolean domain.

%%TC:ignore
\begin{minted}{python}
class ConstraintSat(Constraint):                        
    def __init__(self, elems, weight):
        self._elems = elems       
        self._weight = weight      
    def evaluate (self, variables, domains):  
        res = 1
        
        n = len(self._elems)
        
        for i in range(0,n):
            x = abs(self._elems[i]) - 1   
            if self._elems[i] < 0 :
                res *= (1 - variables[x])
            else:
                res *= variables[x]
        return res * self._weight
    def getWeight(self):
        return self._weight
\end{minted}
%%TC:endignore

The \textbf{binary constraint model} is built around $2$ types of constraints for boolean variables: unary and binary constraints. A unary constraint, will return $0$ if the variable is $false$ and a weight if the variable is $true$. It can be defined as a vector:

%%TC:ignore
$$
\begin{bmatrix}
0 \\
e
\end{bmatrix}
, e \in \mathbb{Z}
$$
%%TC:endignore

The values given should be the variable that is constrained and the weight for the true option.

%%TC:ignore
\begin{minted}{python}
class ConstraintBinaryModelUnary(Constraint):        
    def __init__(self, elem, weight):
        self._elem = elem        	
        self._weight = weight       
        
    def evaluate(self, variables, domains):   
        if variables[self._elem] == 1 :
            return self._weight
        return 0
    def getWeight(self):
        return max(0,self._weight)
\end{minted}
%%TC:endignore

A binary constraint can also be of two types. The first type will assign $0$ if the values are the same, and some (possibly different) weights otherwise. The second will assign $0$ for different values, and some weights for the same.

%%TC:ignore
$$
\begin{bmatrix}
0 & b \\
c & 0
\end{bmatrix}
, b,c \in \mathbb{N}
$$

$$
\begin{bmatrix}
a & 0 \\
0 & d
\end{bmatrix}
, a,d \in \mathbb{N}
$$
%%TC:endignore

The values given should be an array of $2$ variables that are constrained and an array of $2$ values representing the corresponding weights.

%%TC:ignore
\begin{minted}{python}
class ConstraintBinaryModelBinaryDifferent(Constraint):         
    def __init__(self, elems, weights):
        self._elems = elems        	  
        self._weights = weights      
        
    def evaluate(self, variables, domains):    
        if variables[self._elems[0]] == 0 
                and variables[self._elems[1]] == 1 :
            return self._weights[0]
        if variables[self._elems[0]] == 1 
                and variables[self._elems[1]] == 0 :
            return self._weights[1]
        return 0
    def getWeight(self):
        return max(0,np.amax(self._weights))
\end{minted}

\begin{minted}{python}
class ConstraintBinaryModelBinarySame(Constraint):         
    def __init__(self, elems, weights):
        self._elems = elems        	  
        self._weights = weights       
        
    def evaluate(self, variables, domains):    
        if variables[self._elems[0]] == 0 
                and variables[self._elems[1]] == 0 :
            return self._weights[0]
        if variables[self._elems[0]] == 1 
                and variables[self._elems[1]] == 1 :
            return self._weights[1]
        return 0
    def getWeight(self):
        return max(0,np.amax(self._weights))
\end{minted}
%%TC:endignore

The fitness function for the binary constraint model will be the sum of the offset, the weights from the unary constraints and the weights from the binary constraints.

\subsubsection{The \textit{Statistics} class} \label{genStats}

This class is used to compute and plot general statistics, using the \textit{Population} and \textit{Organism} classes to obtain the values for the fitness at each round. From this, we compute the average fitness, the maximum fitness, the minimum fitness, the average fitness of the first $5\%$ organisms by fitness and the average fitness of the last $5\%$ organisms. Those statistics are, at the end, printed on the same graph. All the values shown are normalised by the maximum possible fitness, so our graphs will have values between $0$ and $1$. We consider as the maximum possible fitness the total sum of the weights of all the constraints. In practice, it might be impossible to satisfy all the constraints, so our peak could be below $1$.

%%TC:ignore
\begin{minted}{python}

class Statistics:
    
    def updateStats(self): #updates the stats at each round
        rnd = self._population.getRound()
        avgFit = self._computeAvgFit()
        self._avgFitRaw.append(avgFit)
        self._avgFitList.append((rnd, avgFit / 
            self._maxPossibleFitness))
        self._maxFitList.append((rnd, self._computeMaxFit() / 
            self._maxPossibleFitness))
        self._minFitList.append((rnd, self._computeMinFit() / 
            self._maxPossibleFitness))
        self._max5PerFitList.append((rnd, self._computeMaxPerFit(5) / 
            self._maxPossibleFitness))
        self._min5PerFitList.append((rnd, self._computeMinPerFit(5) / 
            self._maxPossibleFitness))
        
    def plotStats(self):     #plot all the stats so far
        axes = plt.gca()
        axes.set_ylim([0,1])
        plt.title("General fitness statistics")
        plt.xlabel("Rounds")
        plt.ylabel("Normalized fitness")
        self.plotAvgFit()
        self.plotMinFit()
        self.plotMaxFit()
        self.plotMin5PerFit()
        self.plotMax5PerFit()
        self._fillPlot(self._maxFitList, self._max5PerFitList) 
        self._fillPlot(self._minFitList, self._min5PerFitList) 
        plt.legend(bbox_to_anchor=(1.04,1), loc="upper left")
        plt.show()

    def plotMax5PerFit(self):
        self._plotStat(self._max5PerFitList, "Maximum 5% average",
            "green")
        
    def _plotStat(self, statList, des, clr):
        xs = [x[0] for x in statList]
        ys = [x[1] for x in statList]
        plt.plot(xs,ys, label = des, color = clr)
        
    def _computeAvgFit(self):
        s = 0
        orgs = self._population.getOrganisms()
        for i in range(0, self._orgNum):
            s += orgs[i].getFitness()
        return s / self._orgNum
        
    def _computeMinPerFit(self, per):
        per = per / 100
        orgs = self._population.getOrganisms()
        fit = []
        for i in range(0, self._orgNum):
            fit.append(orgs[i].getFitness())
        fit.sort()
        l = int(math.ceil(len(fit) * per))     
        fit = fit[:l]
        s = 0
        for x in fit:                            
            s += x
        return s / l

\end{minted}
%%TC:endignore

\subsubsection{The \textit{LocalStatistics} class} \label{localStatsClass}

This class is used to compute and plot local statistics. Firstly, it plots, for each round, the $\Delta raw fitness$ for the fittest possible mutant, the average of possible mutants and the average of possible mutants that are fitter than their parent, together with the normalisation value.

Secondly, the simulator plots selection coefficient. It uses the fitness data to compute the coefficient for each round. It also computes the standard error, using the propagation of uncertainty.

%%TC:ignore
\begin{minted}{python}
    def _computeError(self, a, b, sa, sb, sab):
        return np.abs(a / b) * np.sqrt((sa / a) * 
            (sa / a) + (sb / b) * (sb / b) - 2 * (sab) / (a * b))

    def _computeSelectionFitterAvg(self, roundNeighbourData):
        rnd = self._population.getRound()
        orgFit = []
        orgAvgFitter = []
        n = 0
        for org in self._population.getOrganisms():
            (fit, mutDFit, maxFit, avgFitter) = 
                roundNeighbourData[org] 
            orgFit.append(fit)
            orgAvgFitter.append(avgFitter)
            n += 1
        sdFit = np.std(orgFit, ddof = 1)            
            #standard deviation of orgFit
        sdFitter = np.std(orgAvgFitter, ddof = 1)   
            #standard deviation of orgAvgFitter
        sdAvgFit = sdFit / (np.sqrt(n))         
            #standard deviation of the mean of orgFit
        sdAvgFitter = sdFitter / (np.sqrt(n))   
            #standard deviation of the mean of of orgAvgFitter
        avgFit = np.mean(orgFit)                    
            #mean of orgFit
        avgFitter = np.mean(orgAvgFitter)           
            #mean of orgAvgFitter
        covFitFitter = np.cov([orgFit,orgAvgFitter], ddof = 1)           
            #covariance between orgFit and orgAvgFitter
        covAvgFitFitter = 1 / (n) * covFitFitter                    
            #covariance between the means                 
        sel = (max(avgFitter - avgFit,0)) / avgFit
        err = self._computeError(avgFitter, avgFit, sdAvgFitter,
            sdAvgFit, covAvgFitFitter[0][1])
        return (sel, err)
\end{minted}
%%TC:endignore

When generating the graph for our selection coefficient, we use the $curve\_fit$ procedure from the $SciPy$ package for fitting an exponential function to show the exponential decay. 

%%TC:ignore
\begin{minted}{python}
#the function used for fitting
def expFunc(x, a, c):
        return a*np.exp(-c*x)

\end{minted}
%%TC:endignore

In order to maximise the fidelity of the curve fitting, we take into account the standard error of the selection coefficient, so that the points with a smaller error have a bigger weight than the others.

%%TC:ignore
\begin{minted}{python}
#this computes the fitted curve
def _compCoeff(self, xs, ys, func, sigma = None):
        popt, pcov = curve_fit(func, xs, ys, sigma = sigma,
            absolute_sigma=True)
        return popt


#the call to the above function with the errors
yerr = [x[1] for x in errorList]
popt = self._compCoeff(xs,ys,func,[max(x,0.001) for x in yerr])

\end{minted}
%%TC:endignore

Finally, this class computes and plots the average and minimum distance to the peak for each round. Thus, if we are given the genome of the peak organism, we compute the distance by going through all the organisms at each round and counting the number of different alleles at the same locus in the two genomes.

%%TC:ignore
\begin{minted}{python}
def _computeDiff(self, a, b):
        n = len(a)
        s = 0
        for i in range(n):
            if a[i] != b[i]:
                s += 1
        return s

#in the method that computes the stats:
for org in self._population.getOrganisms():
        if self._maxGenome != None:
            totalDistToMax += self._computeDiff(
                org.getGenome(), self._maxGenome)
            if diff < minDistToMax:
                minDistToMax = diff
                
#we return
totalDistToMax / numArgs
\end{minted}
%%TC:endignore

\subsection{Appendix E}

We include here the code for implementing the recursive smooth-landscape ($myFunc$), defined in Section \ref{recRes} and used for simulations in Section \ref{semiSmooth}.

%%TC:ignore
\begin{minted}{python}
def myFunc(args):
    
    def neighbours(x):
        tx = tuple(x)
        if tx in cachen:
            return cachen[tx]
        res = []
        for i in range(len(x)):
            #compute all neighbouring mutants
            curr = x.copy()
            curr[i] = 1 - curr[i]
            res.append(curr)
        cachen[tuple(x)] = res
        return res
    
    def s_plus(x):
        tx = tuple(x)
        if tx in cachesp:
            return cachesp[tx]
        fx = myFunc(x)
        mxm = 0
        for y in neighbours(x):
            fy = myFunc(y)
            mxm = max(mxm, fy - fx)
        cachesp[tx] = mxm
        return mxm
    
    def s_minus(x):
        tx = tuple(x)
        if tx in cachesm:
            return cachesm[tx]
        fx = myFunc(x)
        mnm = 1000000000
        spx = s_plus(x)
        for y in neighbours(x):
            fy = myFunc(y)
            if fx + spx > fy and fy > fx:
                mnm = min(mnm, fy - fx)
        if mnm == 1000000000:
            mnm = 0
        cachesm[tx] = mnm
        return mnm
    
    def xor (xs,ys):
        res = []
        for (x,y) in zip(xs,ys):
            if x == y:
                res.append(0)
            else:
                res.append(1)
        return res
    
    def bitfield(n):
        return [int(digit) for digit in bin(n)[2:]] 
            # [2:] to chop off the "0b" part 
    
    m = len(args)
    sp = 1000000000
    sm = 1000000000
    
    for v in range(2 ^ (m - 1), 2 ^ m - 1):
        x = bitfield(v)
        sp = min(sp, s_plus(x))
        sm = min(sm, s_minus(x))
    
    if sm >= sp:
        sm = sp / 2
    
    targs = tuple(args)
    if targs in cache:       
        #if we already computed this function, return the cached value
        return cache[targs]
    if m == 2:
        if args == [0,0]:
            cache[targs] = 2
            return 2
        if args == [0,1]:
            cache[targs] = 3
            return 3
        if args == [1,0]:
            cache[targs] = 4
            return 4
        cache[targs] = 6
        return 6
    a = args[-2]
    b = args[-1]
    x = args[:-2]
    x_star = [0] * (m - 4) + [1,1]
    fx = myFunc(x)
    fx_star = myFunc(x_star)
    if a == 0 and b == 0:
        cache[targs] = fx
        return cache[targs]
    if a == 1 and b == 1:
        cache[targs] = myFunc(xor(x, x_star)) + fx_star + 2 * sp
        return cache[targs]
    if x != x_star:
        cache[targs] = fx + sm
        return cache[targs]
    if a == 0 and b == 1:
        cache[targs] = fx_star + sm
        return cache[targs]
    cache[targs] = fx_star + sp
    return cache[targs]
    
\end{minted}
%%TC:endignore

\newpage

\addcontentsline{toc}{section}{References}
\bibliography{references}

\end{document}